\documentclass[aps,pra,amsmath,reprint,floatfix]{revtex4-1} 

\usepackage{microtype} 
\usepackage{amssymb}
\usepackage{graphicx}
\usepackage{bm}

\RequirePackage[
  hyperindex,colorlinks,bookmarksnumbered,
  plainpages=true,pdfstartview=FitH]{hyperref}
\hypersetup{linkcolor=blue,urlcolor=blue,citecolor=blue}
\usepackage{hyperref}
\usepackage[all]{hypcap}

\renewcommand{\vec}[1]{\bm{#1}}
\newcommand{\ED}{.}
\newcommand{\EC}{,}
\newcommand{\Int}[1]{\int_{#1} \!}
\newcommand{\DimInt}[1]{\int_{#1}' \!}
\newcommand{\DDimInt}[1]{\int_{#1}'' \!}

\newcommand{\Dif}[1]{\! \! \textrm{d}{#1} \ }
\newcommand{\ER}[1]{Eq.~(\ref{#1})}
\newcommand{\EsR}[1]{Eqs.~(\ref{#1})}
\newcommand{\ERn}[1]{(\ref{#1})}
\renewcommand{\FR}[1]{Fig.~\ref{#1}} 
\newcommand{\FsR}[1]{Figs.~\ref{#1}}
\newcommand{\FRn}[1]{\ref{#1}}
\newcommand{\SR}[1]{Sec.~\ref{#1}}

\newcommand{\AR}{\hyperref[appendix]{App.}}

\begin{document}

\title{Fermi-edge singularity and the functional renormalization group}
\author{Fabian B.~Kugler}
\author{Jan von Delft}
\affiliation{Physics Department, Arnold Sommerfeld Center for Theoretical Physics, and Center for NanoScience, Ludwig-Maximilians-Universit\"at M\"unchen, Theresienstr.~37, 80333 Munich, Germany}

\date{\today}

\begin{abstract}
We study the Fermi-edge singularity, describing the response of a
degenerate electron system to optical excitation,
in the framework of the 
functional renormalization group (fRG).
Results for the (interband) particle-hole susceptibility
from various implementations of fRG
(one- and two-particle-irreducible, multi-channel Hubbard-Stratonovich,
flowing susceptibility)
are compared to
the summation of all leading logarithmic (log) diagrams,
achieved by a (first-order) solution of the parquet equations.
For the (zero-dimensional) special case of the X-ray-edge singularity,
we show that the leading log formula
can be analytically reproduced
in a consistent way from 
a truncated, one-loop fRG flow.
However, reviewing the underlying diagrammatic structure,
we show that this derivation relies on fortuitous partial
cancellations special to the form of and accuracy applied to the
X-ray-edge singularity
and does not generalize.
\end{abstract}

\maketitle

\section{Introduction}
Fermi-edge singularities describe infrared divergences
in optical spectra arising from the discontinuity of the electronic distribution.
The advance in the experimental techniques of 
cavity quantum electrodynamics \cite{Gabbay2007, Smolka2014, Sidler2016}
has renewed the need for a precise understanding 
of such response functions of
degenerate Fermi systems to optical excitation.
From a theoretical perspective, the study of the X-ray-edge singularity serves as 
a prototypical fermionic problem
which exhibits a logarithmically divergent 
perturbation theory \cite{Giamarchi2004}.
Whereas a solution of the (interband) particle-hole susceptibility via 
parquet equations \cite{Roulet1969, Nozieres1969} amounts to
rather involved computational effort,
Lange et al.\ \cite{Lange2015} have recently suggested to perform this resummation 
via simple approximations in a functional renormalization group (fRG) scheme.
Here, we confirm that it is, indeed, possible to reproduce the (first-order) 
parquet result from a truncated, one-loop fRG
flow without further approximations.
However, a detailed analysis of the underlying diagrammatic structure shows
that this conclusion relies on fortuitous partial cancellations special 
to the X-ray-edge singularity.
In more detail, experimentally,
X-ray absorption in metals has been a topic of interest for a long time.
Similar measurements with infrared light can be performed using heavily doped semiconductors.
Whereas photon absorption in metals typically excites a localized deep core electron, 
effects due to the mobility of valence-band electrons in semiconductors 
can significantly alter the absorption spectrum \cite{Smolka2014}. 
When a quasi-two-dimensional layer of such a semiconducting material is placed inside an optical cavity,
the reversible light-matter coupling leads to the formation of half-light, half-matter excitations, 
attributed to the so-called polariton \cite{Carusotto2013}.
Properties of the microcavity system are deduced from the polariton, i.e., from the photon dressed
by light-matter interaction, bringing its 
self-energy into focus \cite{Averkiev2007, Baeten2014,*Baeten2015, Pimenov2015,*Pimenov2017}.
To leading order in the coupling,
this self-energy is proportional to the particle-hole susceptibility,
well-known from the standard literature on the 
Fermi-edge singularity \cite{Roulet1969, Nozieres1969, Nozieres1969a, Gavoret1969, Mahan1967a, Mahan1967}.
The effect of light-matter interaction on the photon 
is thus governed by a 
correlation function of the fermionic system.
The basic theoretical formulation of the X-ray-edge singularity 
involves a localized scattering impurity, corresponding 
to a deep core level of a metal.
In this form, the problem is exactly solvable in a one-body approach, 
as performed by Nozi\`eres and De Dominicis \cite{Nozieres1969a}.
This approach is, however, limited to the special case that
the scattering impurity is structureless. 
If the problem is tackled in a many-body treatment, the solution
can be generalized to more complicated
situations and has relevance for other problems involving logarithmic divergences.
This includes the Kondo problem \cite{Abrikosov1965, Fukushima1971}
as well as the generalization to 
scattering processes involving a finite-mass valence-band hole,
as necessary for the description of 
optical absorption in semiconductors \cite{Mahan1967a, Gavoret1969}.
In a diagrammatic treatment of the Fermi-edge singularity,
logarithmic divergences appear at all orders,
demanding resummation procedures.
A suitable resummation,
containing all leading logarithmic (log) diagrams,
can be phrased in terms of parquet equations.
These consist of coupled
Bethe-Salpeter equations
in two-particle channels;
here, distinguished by
antiparallel or parallel conduction-valence-band lines \cite{Roulet1969}.
Parquet equations can be used in a variety of theoretical applications \cite{Bickers2004},
and it is worthwile to explore whether results comparable
or even equivalent to solving those
can be obtained by alternative
resummation techniques, such as fRG. 
The functional renormalization group is a versatile many-body framework,
which has proven to give accurate results for
low-dimensional fermionic systems \cite{Metzner2012, Kopietz2010}.
Different realizations and approximations of an exact hierarchy 
of differential equations for vertex functions allow for
rich resummations in the calculation of correlation functions.
Inspired by Lange et al.\ \cite{Lange2015}, 
we study the Fermi-edge singularity and show that,
for the (zero-dimensional) special case of the X-ray-edge
singularity, it actually is possible 
to analytically derive the (first-order) parquet result
from a one-loop fRG scheme.
However, this derivation relies on fortuitous partial cancellations
of diagrams and cannot be applied to more general situations.
We further show that various truncated fRG flows (see below) do not
provide a full summation of parquet diagrams.
Though this conclusion may seem disappointing,
we believe that the analysis by which it was
arrived at is very instructive
and motivates the extension of \textit{one-loop} fRG by
\textit{multiloop} corrections.
Indeed, in two follow-up 
publications \cite{Kugler2017, Kugler2017a}, 
we present a multiloop fRG flow that does
succeed in summing all parquet diagrams
for generic many-body systems.
The paper is organized as follows.
In \SR{sec:problem}, we give the standard formulation of the Fermi-edge and X-ray-edge singularity.
The basics of the parquet solution are briefly reviewed in \SR{sec:parquet},
before, in \SR{sec:frg}, we introduce the fRG framework in its
one-particle- and two-particle-irreducible form.
In \SR{sec:vertex_flow}, we apply the fRG flow to the fermionic four-point vertex
and construct the particle-hole susceptibility at the end of the flow.
Furthermore, we briefly consider the potential of 
computing this susceptibility using
a Hubbard-Stratonovich transformation.
In \SR{sec:self-energy_flow}, we
rephrase the particle-hole susceptibility as a photonic self-energy
to obtain a ``flowing susceptibility'';
we compare results from using a dynamic and static four-point vertex
and use the latter approach to analytically reproduce the parquet formula.
We also relate our findings to the work by Lange et al.\ \cite{Lange2015}
and show how their treatment can be simplified.
Finally, we present our conclusions in \SR{sec:conclusion}.
\section{Fermi-edge singularity}
\label{sec:problem}
In this section, we review the standard formulation of the Fermi-edge singularity
for a two-band electron system.
We are interested in the (interband) particle-hole susceptibility,
describing the response to optical excitation.
A typical absorption process,
where a photon lifts an electron from the lower to the upper band,
is shown in \FR{fig:disp}(a).
There, we anticipate the simplification to the X-ray-edge singularity,
ignoring kinetic energy in the lower band,
thereby considering a static, photo-excited scattering impurity.
Before going into detail,
let us state more generally 
the Hamiltonian of the Fermi-edge singularity,
\begin{equation}
H' = \sum_{\vec{k}} \epsilon_{\vec{k}}^{\phantom\dag}
c_{\vec{k}}^{\dag} c_{\vec{k}}^{\phantom\dag}
+ \sum_{\vec{k}} E_{\vec{k}}^{\phantom\dag}
d_{\vec{k}}^{\dag} d_{\vec{k}}^{\phantom\dag} 
+ \frac{U}{V} \sum_{\vec{k} \vec{p} \vec{q}} 
c_{\vec{k}+\vec{q}}^{\dag} c_{\vec{k}}^{\phantom\dag} d_{\vec{p}-\vec{q}}^{\dag} d_{\vec{p}}^{\phantom\dag}
\EC
\label{eq:fermi-edge_ham}
\end{equation}
describing a two-band electron system with 
interband (screened) Coulomb interaction of the contact type ($U_{\vec{q}} = U >0$).
The operator $c_{\vec{k}}$ ($d_{\vec{k}}$) annihilates 
an electron in the conduction (valence) band,
$V$ is the volume, and the
dispersion relations $\epsilon_{\vec{k}}$, $E_{\vec{k}}$,
account for any intraband interaction in a Fermi-liquid picture.
This is supposed to work well when electronic energies close to the Fermi level $\mu$,
which we take to be on the order of the conduction-band width, dominate.
Using the effective electron and hole masses, $m$ and $m_h$, one has ($\hbar=1$)
\begin{equation}
\epsilon_{\vec{k}} = \frac{\vec{k}^2}{2m}
\EC
\quad
E_{\vec{k}} = -E_G - \frac{\vec{k}^2}{2m_h}
\EC
\quad
E_G > 0
\ED
\label{eq:disp}
\end{equation}
Note that we further ignore Auger-type interactions containing three $c$ or $d$ operators,
since such transitions are suppressed by the size of the band gap $E_G$.
This allows us to treat electrons from 
both bands as different fermion species,
each with conserved particle number.
With the targeted (leading log) accuracy (cf.~\SR{sec:parquet}),
including spin degeneracy (while keeping the density-density interaction)
only results in a doubled density of states $\rho$ \cite{Nozieres1969a}.
In two space dimensions, the free density of states is
$m / (2\pi)$;
in other cases, one approximates $\rho$ by its value at the Fermi level
[cf.~\ER{eq:gclocal}].
\begin{figure}[!t]
\def\svgwidth{.471\textwidth}
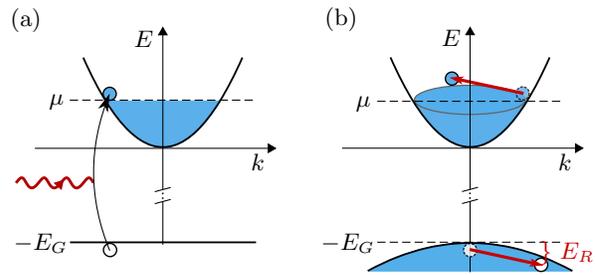
\caption{%
(Color online)
Bandstructure illustrations for two-band electron systems
with chemical potential $\mu$ and band gap $E_G$.
(a) X-ray absorption in metals typically excites a localized, deep core level
to the 
conduction band. 
The flat band acts as a two-level scattering impurity for conduction electrons.
(b) A similar process occurs with infrared light in 
(direct-gap) heavily doped semiconductors.
Only in the limit of 
infinite valence-band (hole) mass,
one reverts to the situation of (a).
Accounting for the mobility of the hole,
scattering processes of conduction electrons on top of the Fermi surface
cost a finite amount of energy, the recoil energy $E_R$.%
}
\label{fig:disp}
\end{figure}
The particle-hole susceptibility
is a two-particle correlation function,
given by
\begin{equation}
i\Pi'(\vec{q}, t) = \frac{1}{V} \sum_{\vec{k}, \vec{p}} \langle \mathcal{T} 
d_{\vec{k}}^{\dag}(t) c_{\vec{k}+\vec{q}}^{\phantom\dag}(t) 
c_{\vec{p}+\vec{q}}^{\dag}(0) d_{\vec{p}}^{\phantom\dag}(0) \rangle
\EC
\end{equation}
with time-ordering operator $\mathcal{T}$.
It exhibits an infrared divergence---%
the Fermi-edge singularity---%
which is cut by the 
(valence-band) recoil energy \cite{Gavoret1969, Pimenov2015,*Pimenov2017} 
at Fermi momentum, equal to $\mu \cdot m/m_h$ 
[cf.~\FR{fig:disp}(b)].
For the case of a polariton experiment using, e.g., a GaAs semiconductor \cite{Smolka2014},
one has a ratio of effective masses between the conduction
and heavy-hole-valence band \cite{Pimenov2015,*Pimenov2017} of $m/m_h \sim 0.14$.
Considering X-ray absorption in metals, one usually encounters the excitation of a localized,
deep core level to the conduction band [cf.~\FR{fig:disp}(a)].
This motivates the severe simplification of an infinite valence-band (hole) mass, 
corresponding to a two-level scattering impurity,
resulting in the Hamiltonian
known from the X-ray-edge singularity, ($\epsilon_d = - E_G < 0$)
\begin{equation}
H = \sum_{\vec{k}} \epsilon_{\vec{k}}^{\phantom\dag} 
c_{\vec{k}}^{\dag} c_{\vec{k}}^{\phantom\dag}
+ \epsilon_d d^{\dag} d 
+ \frac{U}{V} \sum_{\vec{k} \vec{p}} 
c_{\vec{k}}^{\dag} c_{\vec{p}}^{\phantom\dag} d^{\dag} d
\ED
\label{eq:xray_ham}
\end{equation}
Momentum dependencies in interband quantities are completely absorbed 
by the infinitely heavy hole, and only the local conduction-band operators play a role:
\begin{equation}
i\Pi(t) = \langle \mathcal{T} d^{\dag}(t) c(t) c^{\dag}(0) d(0) \rangle
\EC
\quad
c = \frac{1}{\sqrt{V}} \sum_{\vec{k}} c_{\vec{k}}
\ED
\label{eq:particle-hole_susceptibility}
\end{equation}

Without the intrinsic infrared cutoff of the recoil energy,
the (infinite-mass) 
particle-hole susceptibility
shows a true divergence.
In a zero-temperature calculation and
for small interaction,
this takes the form \cite{Roulet1969, Nozieres1969, Nozieres1969a}
\begin{equation}
\Pi(\omega) 
=
\frac{\rho}{2u} \Big[ 1 - \Big( \frac{\omega+\xi_d+i0^+}{-\xi_0} \Big)^{-2u} \Big]
\EC \quad u = \rho U
\ED
\label{eq:irdivergence_phsuscep}
\end{equation}
Here, $-\xi_d = \mu - \epsilon_d = \mu + E_G$
is the threshold frequency and 
$\xi_0 \sim \mu$ an intrinsic ultraviolet cutoff
of the order of the conduction-band width [cf.~\ER{eq:gclocal}].
Note that, for absorption processes, one has an initially 
fully occupied valence band ($E_G \gg k_B T$),
such that $\Pi(t)$ is automatically retarded.
Analogously, the valence-band propagator
$iG^d(t) = \langle \mathcal{T} d(t) d^{\dag} \rangle$
is purely advanced.
Although our calculations will proceed in a finite-temperature formalism,
we aim to reproduce the result \ERn{eq:irdivergence_phsuscep}.
Hence, we numerically consider very low temperatures
and perform the zero-temperature limit in analytic calculations.
As we attribute the constant Hartree part of a fermionic self-energy 
to the renormalized band gap $E_G$,
a diagrammatic expansion using
$G^d(t) \propto \Theta(-t)$ (with the Heaviside step function)
directly shows that 
conduction-band propagators
are not further renormalized by interband interaction.
As already mentioned, the particle-hole susceptibility
can also be viewed as the leading contribution 
(in the light-matter coupling $\rho |M|^2$,
$M$ being the dipole matrix element)
to a photon self-energy.
In the regime under consideration,
electronic processes happen on a timescale $1/\mu$
much shorter than typical times of absorption and emission 
of a photon $1/(\rho |M|^2)$ \cite{Pimenov2015,*Pimenov2017}.
For $\mu \gg \rho |M|^2$, one can thus approximate
the photon self-energy by an interacting particle-hole bubble,
given the standard coupling
\begin{equation}
H'_{\textrm{cpl}} = \frac{1}{\sqrt{V}} \sum_{\vec{p} \vec{q}} \big( 
M 
c_{\vec{p}+\vec{q}}^{\dag} d_{\vec{p}}^{\phantom\dag} a_{\vec{q}}^{\phantom\dag} 
+ 
M^*
d_{\vec{p}}^{\dag} c_{\vec{p}+\vec{q}}^{\phantom\dag} a_{\vec{q}}^{\dag} 
\big)
\EC
\end{equation}
where $a_{\vec{q}}$ annihilates a photon.
For infinite hole mass, the momentum dependence of the
photon absorption can no longer be resolved, and
we use the simplified coupling
\begin{equation}
H_{\textrm{cpl}} = M c^{\dag} d a 
+ 
M^* d^{\dag} c a^{\dag}
\EC
\quad
\sum_{\vec{q}} a_{\vec{q}} = a
\ED
\end{equation}
Having defined the system under consideration [\ER{eq:xray_ham}] and the 
quantity of interest [\ER{eq:particle-hole_susceptibility}],
our analysis will proceed in an imaginary-time action formalism. 
We transform the Grassmann fields for both bands ($c, \bar{c}, d, \bar{d}$) to Matsubara frequencies
according to $c_{\omega} = \int_0^{\beta} \Dif{\tau} c(\tau) e^{i \omega \tau} / \sqrt{\beta}$, etc.,
where $\beta = 1/(k_B T)$.
For the X-ray-edge singularity,
a change to the position basis immediately shows that
conduction-band fields other than the local
ones [cf.~\ER{eq:particle-hole_susceptibility}]
can be integrated out,
leading to the action
\begin{align}
S 
& = - \Int{\omega} G_{0, \omega}^{c,-1} \bar{c}_{\omega} c_{\omega}
- \Int{\omega} G_{0, \omega}^{d,-1} \bar{d}_{\omega} d_{\omega} 
\nonumber 
\\ & \ + 
U \DimInt{\omega \nu \bar{\omega}} \bar{d}_{\omega} d_{\nu} \bar{c}_{\bar{\omega}+\nu} c_{\bar{\omega}+\omega}
\ED
\label{eq:xray_action}
\end{align}
Here, we have introduced a notation where $\Int{\omega}$ is a sum over Matsubara frequencies,
bosonic Matsubara frequencies are denoted by a bar, 
and each prime on an integral sign represents a prefactor of $1/\beta$.
The zero-temperature limit is then conveniently obtained as
\begin{equation}
\lim_{\beta \to \infty}
\DimInt{\omega} f_{\omega}
=
\int \frac{\textrm{d} \omega}{2\pi} f(\omega)
\ED
\label{eq:zerotemp}
\end{equation}
It is worth noting that the action of the more general Fermi-edge singularity,
defined by the Hamiltonian \ERn{eq:fermi-edge_ham},
is perfectly analogous to the one of the X-ray-edge singularity [\ER{eq:xray_action}].
One merely has to identify each Matsubara frequency
with a double index for frequency and momentum $(\omega, \vec{k})$
and Matsubara summations with a double sum
over frequencies and momenta, the prefactor
being $1/(\beta V)$ instead of $1/\beta$.
Hence, all diagrammatic and fRG arguments apply simultaneously to the 
case of finite and infinite hole mass. 
Only for numerical as well as analytic computations,
we restrict ourselves to the 
(zero-dimensional) special case 
of the X-ray-edge singularity,
such that we can readily ignore any momentum dependence.
Whereas for finite hole mass, the propagator
of valence (conduction) electrons is given by
$1/(i\omega + \mu - E_{\vec{k}})$
[$1/(i\omega + \mu - \epsilon_{\vec{k}})$],
for infinite mass,
the valence-band propagator simply reads $G^d_{0, \omega} = 1/(i\omega - \xi_d)$.
As we use a parabolic dispersion in the conduction band, 
we introduce an ultraviolet cutoff $\epsilon_{\vec{k}} \leq \mu+\xi_0$
in momentum space.
The choice of a half-filled conduction band, i.e., $\xi_0 = \mu$, 
yields the particularly simple local propagator
\begin{align}
G^c_{0, \omega} = \frac{1}{V} \sum_{\vec{k}} \frac{1}{i\omega - \epsilon_{\vec{k}} + \mu} =
\rho \int_{-\xi_0}^{\xi_0} \Dif \xi \frac{1}{i\omega-\xi}
\nonumber \\
=
- 2i \rho \arctan( \xi_0 / \omega ) 
\approx - i \pi \rho \, \textrm{sgn} (\omega) \Theta(\xi_0 - |\omega|)
\ED
\label{eq:gclocal}
\end{align}
In the last step, we have ignored any details of the 
ultraviolet cutoff, which are of no physical relevance.
Note that different leading log diagrams typically contain the energy range of
occupied ($\mu$) or unoccupied conduction band states ($\xi_0$) in the argument of the logarithm.
Minor deviations from half-filling, still in the regime of 
$|\mu-\xi_0| \ll \xi_0$, have only subleading effects.
Including photon fields ($a$, $\bar{a}$) into the theory,
one might perform a simple transformation for dimensional reasons 
of the type $\gamma = M a$,
$\bar{\gamma} = M^* \bar{a}$, 
resulting in a rescaled coupling term
\begin{equation}
S_{\textrm{cpl}} = \frac{1}{\sqrt{\beta}} \Int{\bar{\omega} \omega} (
\bar{c}_{\bar{\omega}+\omega} d_{\omega} \gamma_{\bar{\omega}} +
\bar{d}_{\omega} c_{\bar{\omega}+\omega} \bar{\gamma}_{\bar{\omega}} )
\ED
\label{eq:Sgamma}
\end{equation}
Then, in the limit of $M \to 0$, i.e., $G_0^{\gamma} \propto |M|^2 \to 0$, 
one obtains the leading contribution to the photon self-energy $\Pi^{\gamma}$
as precisely the particle-hole susceptibility
\begin{equation}
\lim_{M \to 0} \Pi^{\gamma}_{\bar{\omega}} = \Pi_{\bar{\omega}}
= \DimInt{\omega \nu} \langle \bar{d}_{\omega} d_{\nu} \bar{c}_{\bar{\omega}+\nu} c_{\bar{\omega}+\omega} \rangle
\ED
\label{eq:phsuscep-photon}
\end{equation}
Again, the formula is similarly applicable for the more general
Fermi-edge singularity, where $\bar{\omega}$ denotes frequency and momentum
$(\bar{\omega}, \vec{q})$.
According to the rules of analytic continuation,
$i \bar{\omega} \to \omega+i0^+$,
the X-ray-edge singularity
written in terms of Matsubara frequencies
can directly be inferred from \ER{eq:irdivergence_phsuscep}:
\begin{equation}
\Pi_{\bar{\omega}} 
= 
\frac{\rho}{2u} \Big[ 1 - \Big( \frac{i\bar{\omega}+\xi_d}{-\xi_0} \Big)^{-2u} \Big]
\ED
\label{eq:parquet}
\end{equation}
It is our goal to reproduce this result,
originating from a (first-order) solution of the 
parquet equations, using an fRG scheme.
Before getting into the details of fRG,
let us briefly review the basics of the parquet solution leading
to \ER{eq:parquet}.
\section{First-order parquet solution}
\label{sec:parquet}
We already mentioned
that the X-ray-edge singularity
has been exactly solved
in a one-body approach \cite{Nozieres1969a}
containing the parquet result
\ERn{eq:irdivergence_phsuscep}
in the weak-coupling limit.
For the sake of generalizability to actual fermionic many-body problems,
one is interested in other (approximate) solutions
obtained from a many-body treatment.
Roulet et al.\ \cite{Roulet1969} have achieved such a solution 
of the X-ray-edge singularity
in leading order of the logarithmic singularity.
This \textit{first-order} parquet solution
sums up all perturbative terms of the type
$ u^{n+p} \ln^{n+1} | \xi_0 / (\omega + \xi_d) | $,
where $p=0$.
These correspond to the leading log (or parquet) diagrams;
subleading terms with $p>0$ are neglected.
Such an approximation is applicable for
small interaction, $u \ll 1$,
and frequencies not too close to the threshold $-\xi_d$.
Yet, a subsequent work \cite{Nozieres1969} 
as well as the exact solution \cite{Nozieres1969a}
show that, for small coupling, the result actually holds 
for frequencies arbitrarily close to the threshold.
\begin{figure}[t]
\includegraphics[width=.48\textwidth]{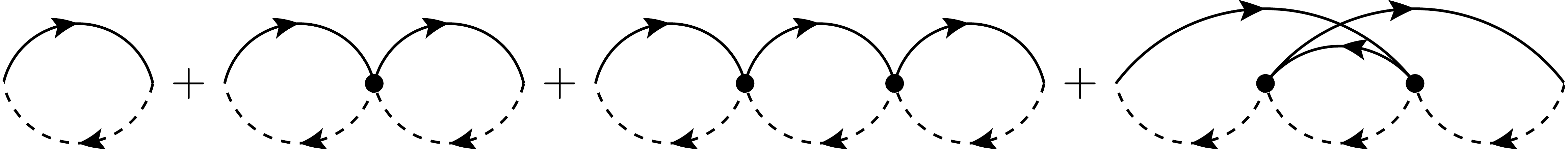}
\caption{%
Particle-hole susceptibility
$\Pi$ [\ER{eq:phsuscep-photon}]
up to second order in the interaction, consisting of
the first three ladder diagrams [L(0), L(1), L(2)]
and the crossed diagram [C(2)].
Full (dashed) lines denote propagators of conduction (valence) electrons. 
Dots represent bare vertices with a factor $-U$.
}
\label{fig:corr2nd}
\end{figure}
The lowest-order diagrams for the particle-hole susceptibility,
corresponding to the first terms of an expansion of \ER{eq:parquet} in $u$,
are shown in \FR{fig:corr2nd}.
Full lines denote conduction-band ($c$)
and dashed lines valence-band ($d$) propagators. 
Self-energy corrections,
affecting the $d$ propagator, can be ignored,
as discussed later.
A bare vertex, symbolized by a solid circle, demands energy(-momentum) conservation 
and multiplication by $-U$. 
Apart from that, there are no combinatorial or sign factors attached to diagrams.  
Free variables are to be integrated over with dimension-full integrals [cf.~\ER{eq:zerotemp}].
The first three diagrams in \FR{fig:corr2nd} are called ladder diagrams.
It is easy to see that taking into account only ladder diagrams 
leads to the false prediction of a bound state \cite{Mahan1967a}.
Crossed diagrams, such as the last diagram in \FR{fig:corr2nd}, are crucial
for an accurate description and encode screening effects (conduction-band holes)
of the Fermi sea.
Figure \FRn{fig:CorrVertex_ParquetOrders}(a) shows how the 
leading log result
is built up in an expansion of \ER{eq:parquet},
exemplified by the real part.
Numerical results in \SR{sec:vertex_flow} and \SR{sec:self-energy_flow}
aim to reproduce this form.
Note that, written in terms of Matsubara frequencies,
the particle-hole susceptibility \ERn{eq:parquet} is no longer singular.
The seemingly quick convergence of the perturbative curves 
to the full solution 
at an interaction parameter $u = 0.28$
in \FR{fig:CorrVertex_ParquetOrders}(a)
is also due to a rapid decay of the 
expansion coefficients.
\begin{figure}[t]
\includegraphics[width=0.235\textwidth]{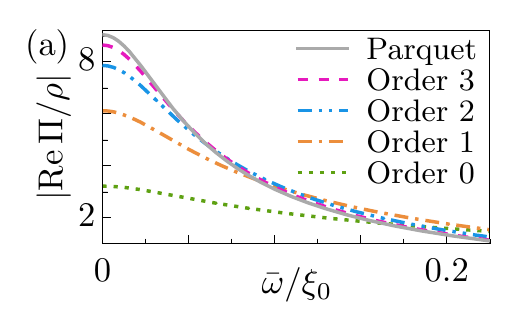}
\quad
\includegraphics[width=0.22\textwidth]{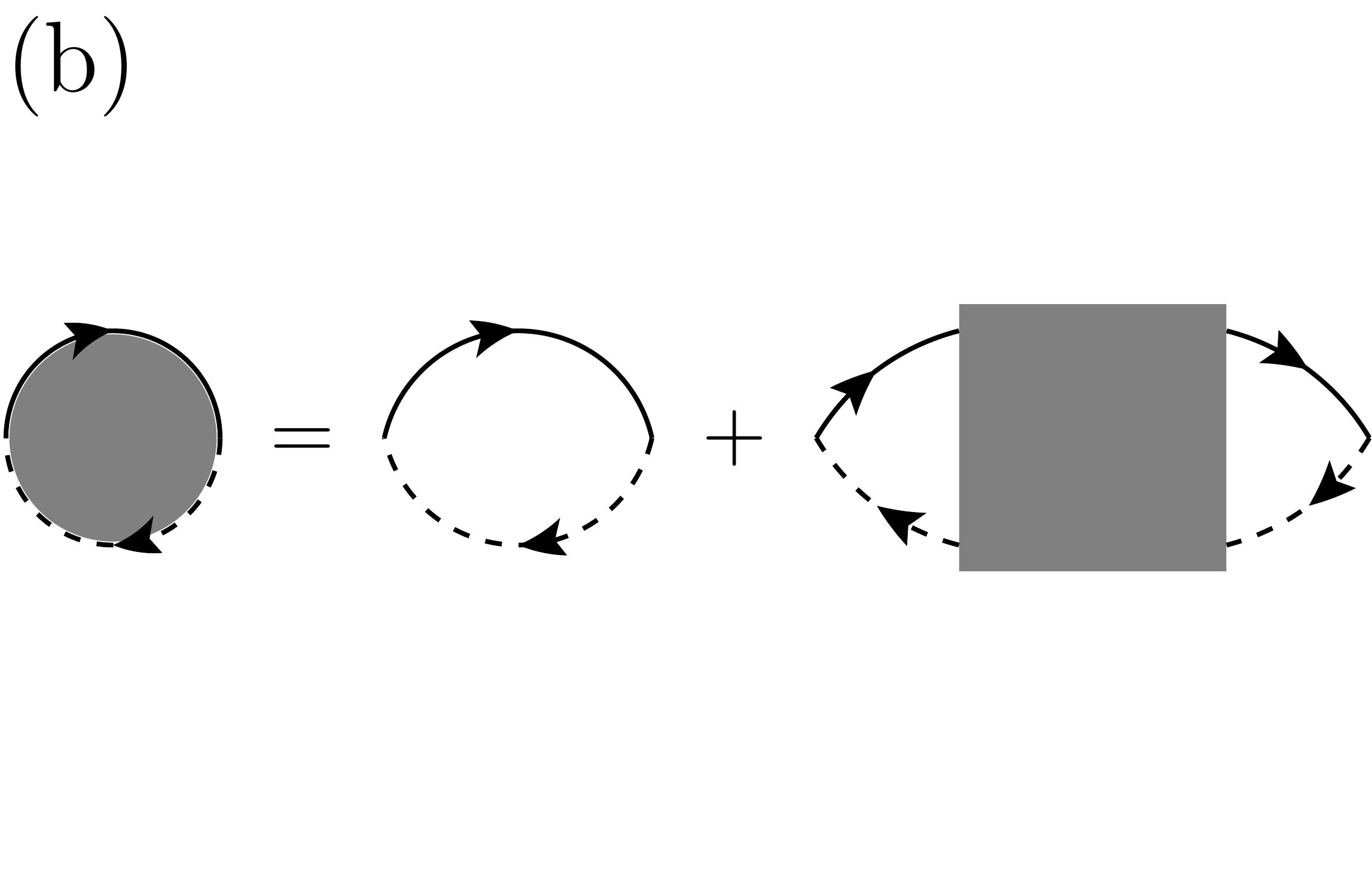}
\caption{%
(a) (Color online)
Leading log formula in terms of Matsubara frequencies [\ER{eq:parquet}] at increasing orders in the coupling $u$.
Numerical parameters are $u = 0.28$, $|\xi_d/\xi_0| = 1/25$, and
the grid for Matsubara frequencies is set by 
$\beta \xi_0 = 500$.
(The same parameters are used throughout this work.)
Here, we show (connected) lines for clarity.
(b) The particle-hole susceptibility $\Pi$ (full circle) can be expressed via 
the bare bubble and the 1PI four-point vertex
$\Gamma^{(4)}$, denoted by a full square,
according to \ER{eq:phsuscep-gamma4}.%
}
\label{fig:CorrVertex_ParquetOrders}
\end{figure}
Though, for real frequencies, $\xi_d$ acts as a frequency shift,
it is a property of the analytic continuation that, 
in imaginary-frequency space, different values for $\xi_d$ stretch/flatten the curve.
Since we have incorporated the physical effect of the size of the band gap
already in the choice of the interaction in the Hamiltonian \ERn{eq:fermi-edge_ham},
we can choose any value for $\xi_d$ in our calculations.
In order to have a pronounced peak in the Matsubara curve,
we take $|\xi_d/\xi_0|=1/25$, implying $u \ln | \xi_0/\xi_d | \approx 0.9$.
Note that, as can be seen from the simple computation of the particle-hole bubble,
zero-temperature calculations are discontinuous w.r.t.\ to $\xi_d$ at
$\xi_d=0$.
Choosing $\xi_d=0$, one loses analytic properties 
and only obtains the real part of the logarithmic factors
depending on $|\bar{\omega}|$ (cf.~\AR).
The four-point correlation function in the particle-hole susceptibility 
can be rephrased by cutting external legs 
(in general, as dressed propagators $G^d$, $G^c$)
in the connected part according to [cf., e.g., Eq.~(6.92) of Ref.~\onlinecite{Kopietz2010}]
\begin{align}
\langle \bar{d}_{\omega} d_{\nu} \bar{c}_{\bar{\omega}+\nu} c_{\bar{\omega}+\omega} \rangle
& = 
G^d_{\omega} G^c_{\bar{\omega} + \omega} \delta_{\omega, \nu}
+
G^d_{\omega} G^d_{\nu} 
\nonumber \\
& \ \times
G^c_{\bar{\omega} + \omega} G^c_{\bar{\omega} + \nu}
\Gamma^{\bar{d} c \bar{c} d}_{\omega, \bar{\omega}+\omega, \bar{\omega}+\nu, \nu} /\beta
\ED
\end{align}
This introduces the one-particle-irreducible (1PI) four-point vertex
$\Gamma^{\bar{d} c \bar{c} d}$.
Consequently, the particle-hole susceptibility is fully
determined by $\Gamma^{(4)} = \Gamma^{\bar{d} c \bar{c} d}$ via
\begin{align}
\Pi_{\bar{\omega}}
& = 
\DimInt{\omega} G^d_{\omega} G^c_{\bar{\omega} + \omega}
+ \DDimInt{\omega \nu} G^d_{\omega} G^d_{\nu} G^c_{\bar{\omega} + \omega} G^c_{\bar{\omega} + \nu} 
\Gamma^{(4)}_{\omega, \bar{\omega}+\omega, \bar{\omega}+\nu, \nu}
\EC
\label{eq:phsuscep-gamma4}
\end{align}
the graphical representation of which is shown in \FR{fig:CorrVertex_ParquetOrders}(b).
The parquet equations are then focused on 
the four-point vertex and use a diagrammatic decomposition
in two-particle channels.
For the Fermi-edge singularity, 
the leading log divergence is determined by the two channels
characterized by parallel and antiparallel conduction-valence-band lines:
\begin{subequations}
\label{eq:parqueteqs}
\begin{align}
\Gamma^{(4)} 
= 
R + \gamma_{p} + & \gamma_{a}
\EC \ \ \,
I_{p} 
= R + \gamma_{a}
\EC 
\ \ \,
I_{a} = R + \gamma_{p}
\EC
\label{eq:parqueteq1}
\\
\gamma_{a;\, \omega, \bar{\omega}+\omega, \bar{\omega}+\nu, \nu} 
& = 
\DimInt{\omega'} I_{a;\, \omega, \bar{\omega}+\omega, \bar{\omega}+\omega', \omega'} 
G^d_{\omega'} G^c_{\bar{\omega}+\omega'}
\nonumber \\ 
& \qquad \times
\Gamma^{(4)}_{\omega', \bar{\omega}+\omega', \bar{\omega}+\nu, \nu}
\EC
\label{eq:parqueteq2}
\\
\gamma_{p;\, \omega, \bar{\nu}-\nu, \bar{\nu}-\omega, \nu} 
& = 
\DimInt{\omega'} I_{p;\, \omega, \bar{\nu}-\omega', \bar{\nu}-\omega, \omega'} 
G^d_{\omega'} G^c_{\bar{\nu}-\omega'}
\nonumber \\ 
& \qquad \times
\Gamma^{(4)}_{\omega', \bar{\nu}-\nu, \bar{\nu}-\omega', \nu}
\ED
\label{eq:parqueteq3}
\end{align}
\end{subequations}
Here, $R$ is the totally (two-particle-) irreducible vertex;
$\gamma_{a}$ and $\gamma_{p}$ 
are reducible
while
$I_{a}$ and $I_{p}$ are
irreducible vertices in the antiparallel and parallel channel, respectively.
Note that a $\Gamma^{(4)}$ diagram can be reducible in 
exclusively one of the two channels \cite{Roulet1969};
diagrams irreducible in both channels belong to $R$.
The Bethe-Salpeter equations for $\gamma_{a}$ \ERn{eq:parqueteq2}
and $\gamma_{p}$ \ERn{eq:parqueteq3},
which are the crucial components
of the parquet equations, are
illustrated in \FR{fig:bethesalpeter}.
\begin{figure}[t]
\includegraphics[width=.48\textwidth]{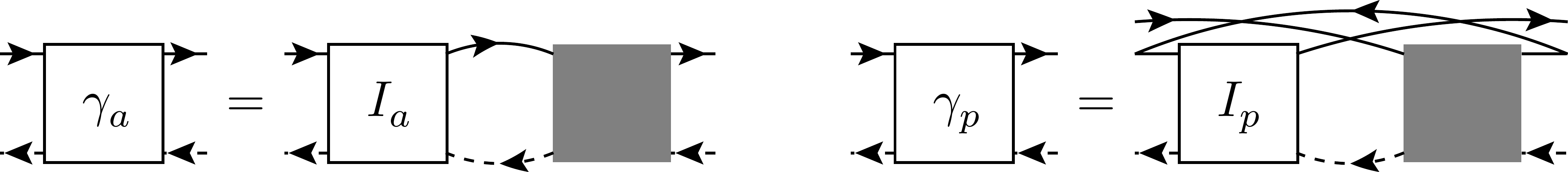}
\caption{%
Bethe-Salpeter equations for both two-particle channels, where
$\gamma_{a}$ and $\gamma_p$ are reducible while
$I_{a}$ and $I_{p}$ are irreducible vertices in antiparallel and
parallel conduction-valence-band lines, respectively. The vertices
are further related via \ER{eq:parqueteq1}.
}
\label{fig:bethesalpeter}
\end{figure}
The parquet equations \ERn{eq:parqueteqs} as such are exact and 
merely represent a classification of diagrams.
In the first-order solution \cite{Roulet1969} 
(also referred to as parquet approximation \cite{Bickers2004}), 
one approximates
the totally irreducible vertex by its bare part,
i.e., $R=-U$.
To be consistent with the leading log summation (of the X-ray-edge singularity),
one further neglects any fermionic self-energies \cite{Roulet1969, Nozieres1969}.
In fact, it is easily shown that
the lowest (non-constant) contribution to $\Sigma^d$
involves the subleading term
$u^2 \ln |\xi_0/(\omega+\xi_d)|$.
Similarly, higher-order corrections to $R$ are subleadingly divergent.
From the exact solution \cite{Nozieres1969a},
it is known that extensions of the first-order parquet scheme just
lead to the replacement of $u$ by more complicated functions of $u$ 
in the characteristic form of the particle-hole susceptibility [\ER{eq:irdivergence_phsuscep}].
For weak coupling, it is thus
justified to focus on the leading-order result.
We will henceforth ignore all fermionic self-energies
and omit the index $0$ 
on fermionic propagators
when referring to the X-ray-edge singularity.
(It should be noted that these arguments do not 
directly apply to \textit{any} Fermi-edge singularity.
In particular, considering a finite-mass valence-band hole,
it was shown that $\Sigma^d$ has a crucial effect on the particle-hole susceptibility
and encodes the influence of indirect transitions \cite{Gavoret1969, Pimenov2015,*Pimenov2017}.)
From the parquet equations \ERn{eq:parqueteqs},
one can also extract the diagrammatic content of the 
emergent four-point vertex $\Gamma^{(4)}$.
All leading log diagrams (parquet graphs) 
are obtained by successively replacing 
bare vertices (starting from the first-order, bare vertex)
by parallel and antiparallel bubbles (cf.~\FR{fig:Parquet3rdOrder}).
Note that such a parquet resummation is 
the natural extension to two channels of
what the ladder summation is to one channel.
Having gained insight into the structure of the parquet equations
and the leading log diagrams,
let us move on to the formalism used in the remainder of this paper.
\begin{figure}[t!]
\includegraphics[width=.28\textwidth]{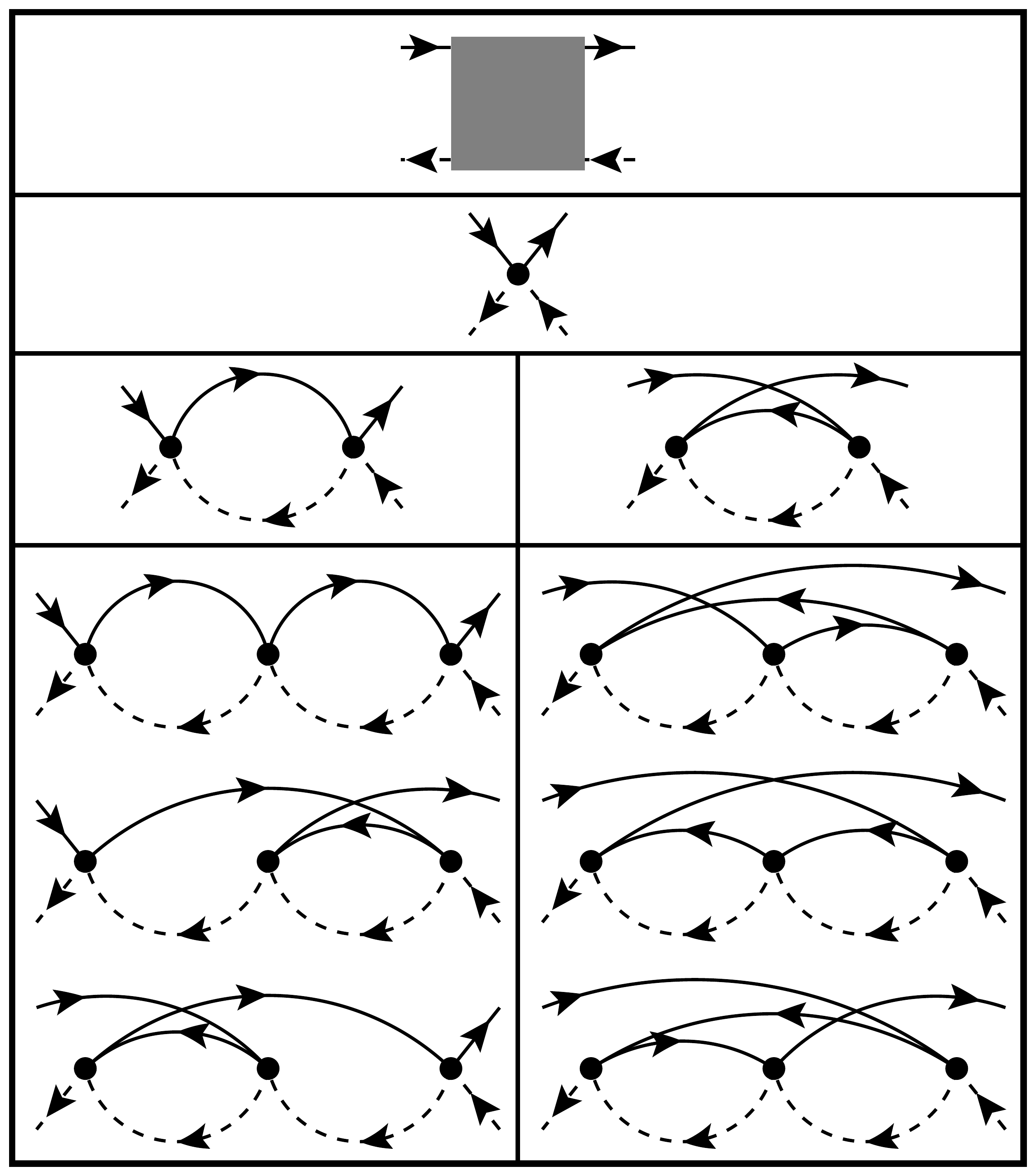}
\caption{%
Parquet graphs for
the four-point vertex
$\Gamma^{(4)}$, consisting of
diagrams reducible in (left) antiparallel lines and 
(right) parallel lines,
up to third order in the interaction.
Note that all diagrams are obtained by 
successively replacing 
bare vertices
by antiparallel and parallel bubbles.%
}
\label{fig:Parquet3rdOrder}
\end{figure}
\section{Functional renormalization group}
\label{sec:frg}
The functional renormalization group (fRG)
is a many-body framework,
which in principle allows one to examine the renormalization group flow
of all coupling constants in their full functional dependence
and to obtain diagrammatic resummations of vertex
and correlation functions. Its basic
idea is to consider the change of a many-body generating functional
upon the variation of an artificially introduced scale parameter,
which can act as an effective infrared cutoff and allows 
to successively integrate out high-energy
degrees of freedom.
This procedure of ``zooming out'' from microscopic to many-body physics,
i.e., the evolution of physical quantities upon 
lowering the scale parameter $\Lambda$,
modulating from a trivial to the full theory (cf.~\FR{fig:frgflow}),
is described by an exact functional differential equation.
\begin{figure}[t]
\def\svgwidth{.5\textwidth}
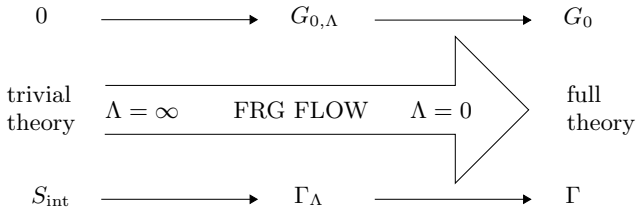
\caption{1PI fRG flow: The flow parameter $\Lambda$, introduced in the quadratic part of the action, 
makes the theory evolve from a trivial to the original, full one. 
At the initial scale, the (quantum) effective action $\Gamma$ can directly be read off from 
the interacting part of the action $S_{\textrm{int}}$. Finally, the desired 
generating functional for 1PI vertices $\Gamma$ is obtained.}
\label{fig:frgflow}
\end{figure}
Most commonly, one incorporates the scale parameter
in the bare propagator of the theory.
Since we are interested in interband quantities such as the particle-hole susceptibility,
it is sufficient to modify the propagator of one band alone.
As $G^d_{0,\omega}$ follows the typical
$1/(i\omega-\xi_d)$ behavior (cf.~\SR{sec:problem}),
it is convenient to choose the lower band.
The appropriate boundary conditions,
to initially ($\Lambda_i=\infty$) extinguish all interband diagrams and finally ($\Lambda_f=0$) 
revert to the original theory,
are $G^d_{0, \Lambda_i}=0$, $G^d_{0, \Lambda_f}=G^d_0$.
We will use two alternative realizations with particularly 
useful computational properties, namely
the $\delta$ regulator,
\begin{align}
G^d_{0, \Lambda,\, \omega} 
& 
= 
\Theta(|\omega|-\Lambda) G^d_{0,\, \omega} 
= 
\frac{\Theta(|\omega|-\Lambda)}{i\omega-\xi_d}
\EC
\nonumber \\
\partial_{\Lambda}
G^d_{0, \Lambda,\, \omega} 
& 
= 
- \delta(|\omega|-\Lambda) G^d_{0,\, \omega} 
=
\frac{- \delta(|\omega|-\Lambda)}{i\omega-\xi_d}
\EC
\label{eq:deltareg}
\end{align}
and the Litim \cite{Litim2001} regulator,
\begin{align}
G^d_{0, \Lambda,\, \omega} 
& 
= 
\frac{1}{i \,\textrm{sgn}(\omega) \,\textrm{max}(|\omega|, \Lambda) -\xi_d}
\EC
\nonumber \\
\partial_{\Lambda}
G^d_{0, \Lambda,\, \omega} 
&
= 
\frac{-i \,\textrm{sgn}(\omega) \Theta(\Lambda-|\omega|)}{[ i \,\textrm{sgn}(\omega) \Lambda - \xi_d ]^2}
\ED
\label{eq:Lreg}
\end{align}
In an exact solution of the flow,
all regulators give identical results
since, at the end of the flow ($\Lambda_f=0$),
the original theory is restored.
However, once approximations are made, the outcomes
might differ significantly.
In particular, this will happen
once the flow of certain quantities 
does not form a total derivative of diagrams, e.g.,
due to truncation.
One can consider different functionals paraphrasing the many-body problem
under the fRG flow.
Two common choices are the 
(quantum) effective action and
the Luttinger-Ward functional serving as generating functionals
for one-particle-irreducible (1PI) and two-particle-irreducible (2PI)
vertices, respectively.
Our study is focused on 1PI fRG flows.
We will only briefly mention the 2PI formulation
to show that this provides no benefit for our treatment.
\subsection{One-particle-irreducible formulation}
The (quantum) effective action $\Gamma$ is obtained from 
the (log of the) partition function---%
in the presence of sources coupled directly to the fields
($S_{\textrm{src}} = \int_{\alpha} j_{\alpha} \varphi_{\alpha}$)---%
by a Legendre transformation.
Its behavior under the flow is given by the 
(so-called) Wetterich equation \cite{Wetterich1993}.
In the notation of Ref.~\onlinecite{Kopietz2010},
particularly useful for mixed (fermionic and bosonic) theories,
it is stated as
\begin{align}
\partial_{\Lambda} \Gamma_{\Lambda}[\bar{\varphi}]
& = 
- \frac{1}{2} \textrm{STr} \Bigg\{
\Big(
 \partial_{\Lambda }G_{0, \Lambda}^{-1} 
\Big) 
\nonumber \\
& \ \times
 \Bigg(
 \bigg[
 \bigg(
 \frac{\delta^2 \Gamma_{\Lambda}[\bar{\varphi}]}{\delta \bar{\varphi} \delta \bar{\varphi}} 
 \bigg)^{\textrm{T}}
 - G_{0, \Lambda}^{-1}
 \bigg]^{-1}
 + G_{0, \Lambda}
 \Bigg)
 \Bigg\}
\ED
\label{eq:gammaflow}
\end{align}
Here, the super trace runs over multi-indices $\alpha$, which specify 
field as well as conjugation indices 
and all further quantum numbers, 
and contains a minus sign when summing over fermionic degrees of freedom.
If the propagator of all fields is set to zero at the beginning of the flow,
the initial condition for $\Gamma$ is given by the interacting part of the action \cite{Kopietz2010},
$\Gamma_{\Lambda_i} = S_{\textrm{int}}$
(no renormalization of vertices by propagating degrees of freedom is possible).
Although we choose only the bare valence-band propagator to be $\Lambda$-dependent,
all interband quantities are still given by the bare interactions of $S_{\textrm{int}}$.
In order to tackle the fundamental and in general unsolvable flow equation \ERn{eq:gammaflow},
$\Gamma$ can be expanded in terms
of 1PI $n$-point vertices $\Gamma^{(n)}$, where we set
\begin{equation}
\Gamma^{(n)}_{\alpha_1 \dots \alpha_n} = \beta^{\frac{n}{2}-1}
\frac{\delta^n \Gamma[\bar{\varphi}]}{\delta \bar{\varphi}_{\alpha_1} \dots \delta \bar{\varphi}_{\alpha_n}}
\bigg|_{\bar{\varphi}=0}
\ED
\label{eq:1PIvertices}
\end{equation}
The functional differential equation \ERn{eq:gammaflow} is transformed into a hierarchy of
infinitely many coupled ordinary differential equations with an interesting structure \cite{Kopietz2010}:
$\partial_{\Lambda} \Gamma^{(n)}$ depends on other vertices only up to $\Gamma^{(n+2)}$
and, then, always via $\textrm{STr} \{ \Gamma^{(n+2)} S \}$.
Here, $S$ is the (so-called) single-scale propagator
$S = - G ( \partial_{\Lambda} G_0^{-1} ) G$,
adding self-energy corrections to a differentiated bare line.
Since, with logarithmic accuracy (cf.~\SR{sec:parquet}), 
we can neglect fermionic self-energies,
we have the notable simplification $S = \partial_{\Lambda} G_0$.
The most common truncation of the still unsolvable hierarchy of flow equations
is to leave higher-order vertices constant 
($\Gamma^{n > n_0}_{\Lambda} = \Gamma^{n > n_0}_{\Lambda_i}$)
yielding a finite set of differential equations.
This has a weak coupling motivation,
as higher-order vertices typically are of increasing order in the interaction.
Furthermore, for a four-point interaction as in our fermionic theory,
the only non-zero initial condition
of a 1PI interband vertex is 
$\Gamma^{\bar{d} c \bar{c} d} = -U$.
Note that, when specifying a vertex, 
we usually omit the superscript $(n)$ and, instead,
write field indices as superscripts and quantum numbers as indices.
With the photon included in the theory, we have the additional non-trivial initial condition
$\Gamma^{\bar{c} d \gamma}_{\Lambda_i,\, \omega, \omega-\bar{\omega}, \bar{\omega}} 
= 1
= \Gamma^{\bar{d} c \bar{\gamma}}_{\Lambda_i,\, \omega-\bar{\omega}, \omega, \bar{\omega}} $
for the mixed three-point vertex.
The flow equations of the individual vertices
are obtained by performing the vertex expansion \ERn{eq:1PIvertices}
on both sides of the Wetterich equation \ERn{eq:gammaflow}.
Given a certain truncation and
the above mentioned initial conditions,
the set of differential equations can be solved by standard methods,
possibly requiring further approximations.
Solutions for the self-energy [$\Gamma^{(2)}$] 
or higher-order vertex functions [$\Gamma^{(n>2)}$] 
can be used to compute correlation functions,
such as the particle-hole susceptibility [cf.~\ER{eq:phsuscep-gamma4}].
For future reference, let us already state
the 1PI fRG flow equation for the four-point vertex
in the purely fermionic theory
[in the matrix notation of \ER{eq:gammaflow}, 
we omit the second index for one-particle quantities:
$G^{c \bar{c}}_{\omega, \omega} = G^c_{\omega}$, etc.].
To describe the leading logarithmic divergence of the Fermi-edge singularity,
we only consider interband combinations of four-point vertices and obtain
\begin{align}
\partial_{\Lambda} 
&
\Gamma^{\bar{d} c \bar{c} d}_{\Lambda,\, \omega, \bar{\omega}+\omega, \bar{\omega}+\nu, \nu} 
%
= 
\DimInt{\omega'} S^d_{\Lambda,\, \omega'} 
\nonumber \\ & 
\!\!\!\! \times \Big(
\Gamma^{\bar{d} c \bar{c} d}_{\Lambda,\, \omega, \bar{\omega}+\omega, \bar{\omega}+\omega', \omega'}
G^c_{\bar{\omega}+\omega'}
%
%
\Gamma^{\bar{d} c \bar{c} d}_{\Lambda,\, \omega', \bar{\omega}+\omega', \bar{\omega}+\nu, \nu}
\nonumber \\ & \ 
+
%
\Gamma^{\bar{d} c \bar{c} d}_{\Lambda,\, \omega, \bar{\nu}-\omega', \bar{\nu}-\omega, \omega'}
G^c_{\bar{\nu}-\omega'}
%
\Gamma^{\bar{d} c \bar{c} d}_{\Lambda,\, \omega', \bar{\nu}-\nu, \bar{\nu}-\omega', \nu}
\nonumber \\ & \
+
\Gamma^{\bar{d} c \bar{c} d \bar{d} d}_%
{\Lambda,\, \omega, \bar{\omega}+\omega, \bar{\omega}+\nu, \nu, \omega', \omega'} 
\Big)
\EC
\quad
\bar{\nu} = \bar{\omega}+\omega+\nu
\ED
\label{eq:gamma4flow}
\end{align}
\begin{figure}[t]
\includegraphics[width=0.484\textwidth]{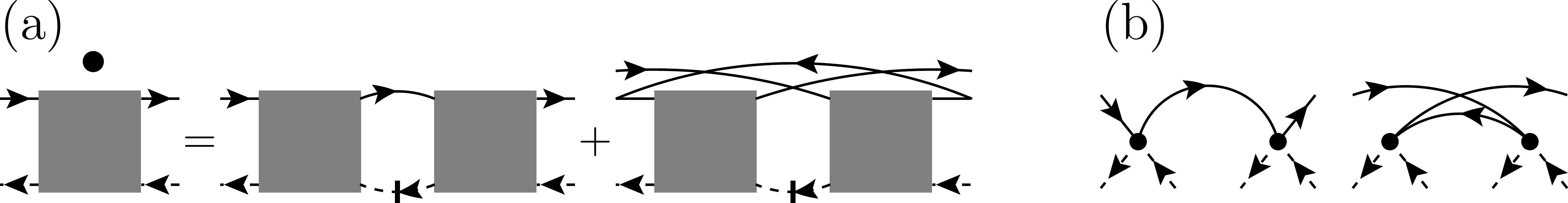}
\caption{%
(a)
Diagrammatic representation of the flow equation \ERn{eq:gamma4flow} 
for $\Gamma^{\bar{d} c \bar{c} d}$
upon neglecting the six-point vertex.
The dot denotes the differentiated vertex;
lines with a vertical dash symbolize the single-scale propagator.
(b)
Three-particle vertices 
$\Phi^{\bar{d} \bar{c} d c d \bar{d}}$ and
$\Phi^{\bar{d} c \bar{c} d d \bar{d}}$,
responsible for the 2PI fRG flow of $I_{p}$ and $I_{a}$, respectively,
at second order in $U$.%
}
\label{fig:gamma4flow}
\end{figure}
Without fermionic self-energies, the propagators 
$G^c$, $G^d$, and $S^d$ are known functions.
If the fRG hierarchy is further truncated by 
discarding the six-point vertex,
$ \Gamma^{(6)}_{\Lambda} = \Gamma^{(6)}_{\Lambda_i} = 0$,
the resulting flow equation is closed in itself 
and can be solved as such.
Figure \FRn{fig:gamma4flow}(a) illustrates this flow equation,
where we denote a single-scale propagator, i.e., a differentiated $d$ line,
by a vertical dash next to the arrow.
Evidently, the 1PI fRG scheme does not yield separate 
flow equations for four-point vertices distinguished
in two-particle channels, in contrast
to the parquet equations \ERn{eq:parqueteqs}.
However, one immediately sees in \FR{fig:gamma4flow}(a)
that contributions from the first summand 
are reducible in antiparallel lines,
whereas contributions from the second one are reducible in parallel lines.
Totally irreducible
diagrams are still present in \ER{eq:gamma4flow} as initial condition (the bare vertex)
and encoded in $\Gamma^{(6)}$,
but, importantly, contributions
from $\textrm{STr}\{ \Gamma^{(6)}S \}$
are also relevant for higher-order parquet diagrams 
in both channels (cf.~\SR{sec:vertex_flow}).
To explore the possibility of treating the
two-particle channels separately from the outset,
let us sketch the 
applicability of 2PI fRG to the Fermi-edge singularity.
\subsection{Two-particle-irreducible formulation}
The 2PI formulation of fRG is based on the Luttinger-Ward functional $\Phi$,
obtained by a Legendre transformation from the (log of the) partition function 
with sources coupled to two fields
($S_{\textrm{src}} = \int_{\alpha \alpha'} \varphi_{\alpha} J_{\alpha \alpha'} \varphi_{\alpha'}$).
It can be shown \cite{Rentrop2015} and is intuitive from its diagrammatic expansion that,
contrary to $\Gamma$,
$\Phi$ does not explicitly depend on the bare propagator of the theory.
The scale dependence is only given by its argument $\mathcal{G}$, representing the full propagator.
Therefore, 
one immediately derives the flow equations
\begin{subequations}
\begin{align}
\partial_{\Lambda} \Phi[\mathcal{G}] & = 
\frac{1}{2} \textrm{STr} \Big\{ 
\frac{\delta \Phi}{\delta \mathcal{G}} \partial_{\Lambda} \mathcal{G} \Big\}
\EC
\label{eq:2piPhiflow} \\
\partial_{\Lambda} \Phi^{(2n)}_{\Lambda, \alpha^{\phantom'}_1 \alpha'_1 \dots \alpha^{\phantom'}_n \alpha'_n}
& =
\frac{1}{2\beta} 
\sum_{\tilde{\alpha}, \tilde{\alpha}'}
\Phi^{(2n+2)}_{\Lambda, \alpha^{\phantom'}_1 \alpha'_1 \dots \alpha^{\phantom'}_n \alpha'_n \tilde{\alpha} \tilde{\alpha}'}
\partial^{\phantom\dag}_{\Lambda} G^{\phantom\dag}_{\tilde{\alpha} \tilde{\alpha}'}
\EC
\label{eq:2piflow}
\end{align}
\end{subequations}
where $G$ is the physical propagator $\mathcal{G}|_{J=0}$. 
Equation \ERn{eq:2piPhiflow} has
a much simpler structure compared 
to the Wetterich equation \ERn{eq:gammaflow}.
The 2PI $n$-particle vertices,
as coefficients of $\Phi$ when expanded around the
physical propagator,
\begin{align}
\Phi^{(2n)}_{\alpha^{\phantom'}_1 \alpha'_1 \dots \alpha^{\phantom'}_n \alpha'_n} & = 
\beta^{n-1}
\frac{\delta^n \Phi}{\delta \mathcal{G}_{\alpha_1 \alpha'_1} \dots \delta \mathcal{G}_{\alpha_n \alpha'_n}} 
\bigg|_{\mathcal{G}=G}
\EC
\end{align}
are primarily suited (to compute correlation functions)
for a purely fermionic theory,
where vertices only connect an even number of fields. 
Unlike the totally antisymmetric 1PI four-point vertex
(where particularly $\Gamma^{\bar{d} c \bar{c} d} = \Gamma^{\bar{d} \bar{c} d c}$),
we have
$\Phi^{\bar{d} \bar{c} d c} = I_{p}$
and
$\Phi^{\bar{d} c \bar{c} d} = I_{a}$, 
implying the desired distinction between the two-particle channels.
(Note that the parquet approximation, which considers only the bare vertex as the
totally irreducible contribution in $I_{p}$ and $I_{a}$ has not yet been made.)
In contrast to the parquet equations,
the 2PI flow, however, does not interrelate these 
two-particle vertices;
instead, it demands the computation of corresponding three-particle vertices.
Moreover, since the 2PI vertices $\Phi^{(2n)}$ are not necessarily 1PI, their
initial conditions are more complex than those of $\Gamma^{(n)}$:
We have $\Phi_{\Lambda_i}^{(2n)} \neq 0$ for infinitely many $n$,
namely for all $\Phi^{(2n)}$ which contain diagrams without internal 
valence-band lines [cf.\ \FR{fig:gamma4flow}(b)].
Therefore,
truncation schemes need to be devised more carefully in the 2PI formulation.
The flow equations for $I_{p}$ and $I_{a}$, deduced from \ER{eq:2piflow},
\begin{subequations}
\label{eq:2PIfloweqs}
\begin{align}
\partial_{\Lambda} I_{p;\, \omega_1, \omega_2, \omega_3, \omega_4} 
& = 
\DimInt{\omega} 
\Phi^{\bar{d} \bar{c} d c d \bar{d}}_{\omega_1, \omega_2, \omega_3, \omega_4, \omega, \omega} \partial_{\Lambda} G^d_{\omega}
\EC \\
\partial_{\Lambda} I_{a;\, \omega_1, \omega_2, \omega_3, \omega_4} 
& = 
\DimInt{\omega} 
\Phi^{\bar{d} c \bar{c} d d \bar{d}}_{\omega_1, \omega_2, \omega_3, \omega_4, \omega, \omega} \partial_{\Lambda} G^d_{\omega}
\EC
\end{align}
\end{subequations}
require knowledge about six-point vertices,
for which an exact consideration is
numerically out of reach (similar to $\Gamma^{(6)}$).
The lowest-order diagrams of 
$\Phi^{\bar{d} \bar{c} d c d \bar{d}}$ and
$\Phi^{\bar{d} c \bar{c} d d \bar{d}}$
are depicted in \FR{fig:gamma4flow}(b).
The simplest way of generating a non-perturbative flow 
is to replace
bare vertices with interacting four-point vertices,
which are then part of the flow.
As opposed to previous proposals,
namely to replace the bare interaction
$U_{\alpha_1 \alpha_2 \alpha_3 \alpha_4}$
by $\Phi^{(4)}_{\alpha_1 \alpha_2 \alpha_3 \alpha_4}$ \cite{Dupuis2005}
or by an average over $\Phi^{(4)}$ with different index permutations \cite{Rentrop2015},
we suggest that the diagrammatically most sensible choice is the 1PI four-point vertex.
Here, this amounts to replacing
$ -U$ by $
\Gamma^{\bar{d} c \bar{c} d}
= 
\Phi^{\bar{d} \bar{c} d c}
+
\Phi^{\bar{d} c \bar{c} d}
-R$ [cf.~\ER{eq:parqueteq1}].
The 1PI four-point vertex $\Gamma^{\bar{d} c \bar{c} d}$
incorporates all possible diagrams;
since both 2PI vertices contain the totally irreducible vertex $R$, 
it must be subtracted.
$\Gamma^{\bar{d} c \bar{c} d}$ also has the full crossing (index-permutation) symmetry
as the bare interaction. Overcounting does not occur
since both vertices are separated by an open $d$ line
and connecting $\partial_{\Lambda} G$ to this approximation of
$\Phi^{\bar{d} \bar{c} d c d \bar{d}}$ and
$\Phi^{\bar{d} c \bar{c} d d \bar{d}}$
induces diagrams reducible in antiparallel and parallel lines, respectively.
Since no further totally irreducible diagram for the 2PI vertices
on top of the initial condition will be generated,
it is consistent to use $R=-U$ 
in the relation for $\Gamma^{\bar{d} c \bar{c} d}$ [\ER{eq:parqueteq1}].
It is possible to evolve $I_{p}$ and $I_{a}$ separately,
using the above described approximations in \ER{eq:2PIfloweqs},
and check the consistence with the parquet equations \ERn{eq:parqueteqs},
interrelating both of them, during the flow.
However, in the ultimately interesting combination 
[cf.~\ER{eq:phsuscep-gamma4} and \ERn{eq:parqueteqs}],
one has the flow
$\partial_{\Lambda} \Gamma^{\bar{d} c \bar{c} d} =
\partial_{\Lambda} I_p +
\partial_{\Lambda} I_a$.
Combining the diagrams of \FR{fig:gamma4flow}(b) with full vertices
and attaching the scale-derived propagator
(here, equal to the single-scale propagator),
we find exactly the same flow equation for the four-point vertex as
given in the truncated 1PI system [\FR{fig:gamma4flow}(a)].
The replacement of $S_{\Lambda}$ by
$\partial_{\Lambda} G$ in the flow of the four-point vertex
when neglecting the six-point vertex,
which is very natural in the above prescription,
is a well known correction \cite{Metzner2012}
that has been found to  
lead to smaller errors in Ward identities \cite{Katanin2004}.
Finally,
we conclude that the above simple 2PI fRG flow
does not enrich the possibilities for an fRG treatment of the Fermi-edge singularity
compared to the 1PI framework.
\section{Correlator from evolved vertices}
\label{sec:vertex_flow}

In this section,
we start to present the results of
our fRG treatment of the X-ray-edge singularity.
First, we perform the fRG flow of vertices
and construct the particle-hole susceptibility
at the end of the flow.
More precisely,
we examine the flow equation \ERn{eq:gamma4flow}
in more detail and compare the resulting form 
of the particle-hole susceptibility,
obtained from the relation \ERn{eq:phsuscep-gamma4},
with the leading log result \ERn{eq:parquet}.
We briefly check whether
it is useful to perform a (multi-channel) Hubbard-Stratonovich transformation
to generate parquet diagrams in the
particle-hole susceptibility 
from combining several 1PI vertices,
finding that this is not the case.
\subsection{Fermionic four-point vertex}
According to \ER{eq:phsuscep-gamma4},
the fermionic four-point vertex is sufficient
to compute the particle-hole susceptibility.
In \ER{eq:gamma4flow}, we have already given its flow equation.
Since a vertex with more than four arguments (and a meaningful resolution in frequency space)
is numerically intractable, we neglect the six-point vertex by truncation 
and obtain the simplified flow for $\Gamma^{(4)}$
illustrated in \FR{fig:gamma4flow}(a).
Solving this flow equation numerically
with the initial condition
$\Gamma^{(4)}_{\Lambda_i} = -U$,
the final form of the particle-hole susceptibility 
[using \ER{eq:phsuscep-gamma4}]
is shown in \FR{fig:Corr4}(a).
We find overall qualitative agreement between both the numerical
and the analytic curve.
Quantitatively, there are disagreements to the leading log
result depending on the choice of regulator,
which originate from neglecting $\Gamma^{(6)}$ in 
the flow of \FR{fig:gamma4flow}(a).
The reason why the $\delta$ regulator yields much
better results than the Litim regulator has recently
been clarified in Ref.~\onlinecite{Kugler2017}: 
The former gives less weight to multiloop corrections 
that are neglected in the present approach.
\begin{figure}[t]
\includegraphics[width=.48\textwidth]{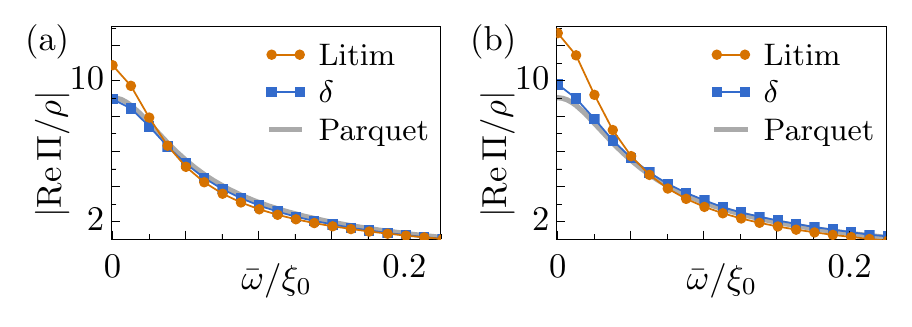}
\caption{%
(Color online)
(a)
Particle-hole susceptibility $\Pi$
computed via $\Gamma^{(4)}$ [\ER{eq:phsuscep-gamma4}], 
which is obtained from a numerical solution of the
truncated flow 
[cf.~\FR{fig:gamma4flow}(a)].
Different results are generated using a Litim or $\delta$ regulator [cf.~\EsR{eq:deltareg}, \ERn{eq:Lreg}]
and compared to the leading log formula \ERn{eq:parquet}.
(b)
$\Pi$ obtained from a numerical solution of the 
flow in the light-matter system [\FsR{fig:gamma4flow}(a), \FRn{fig:photon_flow}].
Stronger deviations (for both regulators) from the parquet curve compared to (a)
occur since the truncated photon flow neglects derivatives of parallel bubbles.%
}
\label{fig:Corr4}
\end{figure}
Let us briefly indicate which types of differentiated diagrams
are missing in the flow equation when neglecting $\Gamma^{(6)}$:
One can easily check, by inserting the second-order diagrams
of $\Gamma^{(4)}$ (cf.~\FR{fig:Parquet3rdOrder})
on the l.h.s.\ and the bare vertex on the r.h.s.,
that the truncated
flow equation [\FR{fig:gamma4flow}(a)]
is satisfied at second order in the interaction.
Note that (without fermionic self-energies) a diagram is simply differentiated
by summing up all copies of this diagram in which one $d$ line is replaced
by a single-scale propagator $S=\partial_{\Lambda}G_0$
at any position (product rule).
At third order, however, the simplified flow equation is no longer fulfilled
since the six-point vertex [neglected in \FR{fig:gamma4flow}(a)]
starts contributing. 
Indeed, the four terms coming from $\textrm{STr} \{ \Gamma^{(6)} S \}$,
depicted in \FR{fig:gamma6contr} (but neglected in the present scheme),
generate the remaining derivatives of third-order parquet
diagrams (cf.~\FR{fig:Parquet3rdOrder}).
We emphasize that all (differentiated) diagrams generated 
by the truncated flow [\FR{fig:gamma4flow}(a)] are of the parquet type.
Indeed, totally (two-particle-) irreducible diagrams of $\Gamma^{(4)}$
exceeding the bare vertex
[corresponding to higher-order contributions of $R$ in the parquet equations \ERn{eq:parqueteqs}]
require proper inclusion of the six-point vertex
(and intraband four-point vertices).
Similar to the recipe given in \SR{sec:parquet},
the truncated flow builds on the bare vertex
by incorporating antiparallel and parallel bubbles
and therefore only generates parquet graphs.
Within the class of leading log diagrams,
the six-point vertex is needed to provide \textit{all} derivatives
of diagrams of $\Gamma^{(4)}$, starting at third order in $U$ (cf.~\FR{fig:gamma6contr}).
In fact, it is easy to see that, in the fRG hierarchy, the parquet graphs comprise
(1PI as well as 2PI) $n$-point vertices of arbitrarily large $n$:
Cutting a valence-band line (without leaving a single conduction-band line in the case of a 1PI description)
generates a vertex of order two higher without leaving the class of parquet graphs.
The corresponding higher-point vertices are required in the flow
via the universal contribution
$\textrm{STr} \{ \Gamma^{(n+2)}_{\Lambda} S_{\Lambda} \}$
or
$\textrm{STr} \{ \Phi^{(n+2)}_{\Lambda} \partial_{\Lambda} G_{\Lambda} \}$
[cf.~\EsR{eq:gammaflow}, \ERn{eq:2piflow}].
Simply truncating the (purely fermionic) fRG hierarchy of flow equations will thus always dismiss contributions to parquet graphs.
\begin{figure}[t]
\includegraphics[width=.48\textwidth]{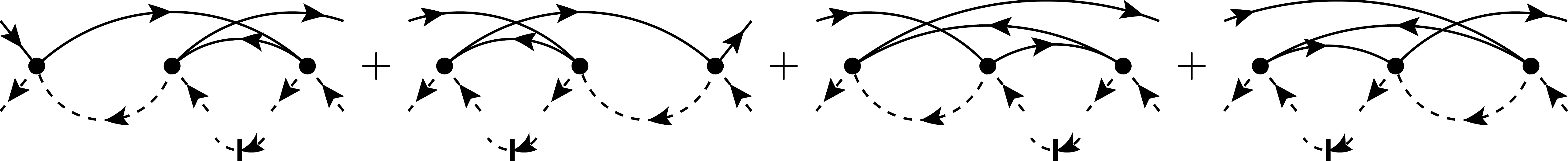}
\caption{%
Third-order contributions from the six-point vertex 
to the flow of $\Gamma^{(4)}$
via $\textrm{STr} \{ \Gamma^{(6)} S \}$,
neglected by the truncated flow in \FR{fig:gamma4flow}(a).
($S$ is graphically separated for clarity.)%
}
\label{fig:gamma6contr}
\end{figure}
The question of how to sum up all parquet diagrams 
in the fermionic four-point vertex via fRG
is beyond the scope of the present work and
is addressed in Ref.~\onlinecite{Kugler2017} 
using a multiloop flow.
Here, instead, we explore various other ways of computing
$\Pi_{\bar{\omega}}$ by using one-loop fRG,
proceeding with auxiliary bosonic fields.
\subsection{Hubbard-Stratonovich fields}
Hubbard-Stratonovich (HS) transformations are used
in the context of several approximation techniques in many-body problems.
Such an exact transformation reformulates the fermionic two-particle interaction 
in terms of propagating auxiliary particles.
For instance,
the lowest-order contribution to a bosonic self-energy
already encodes a ladder summation in the corresponding susceptibility. 
For a parquet resummation, it seems therefore sensible to
perform a multi-channel HS transformation \cite{Lange2015}.
With bosonic fields for the exchange ($\chi$) and pairing ($\psi$) channels,
one has the identification
\begin{align}
S_{\textrm{HS}} 
& = 
\Int{\bar{\omega}} U_{\chi}^{-1} \bar{\chi}_{\bar{\omega}} \chi_{\bar{\omega}}
+
\frac{1}{\sqrt{\beta}} \Int{\bar{\omega} \omega} 
\big(
\bar{c}_{\bar{\omega}+\omega} d_{\omega} \chi_{\bar{\omega}} +
\bar{d}_{\omega} c_{\bar{\omega}+\omega} \bar{\chi}_{\bar{\omega}}
\big)
\nonumber \\
& +
\Int{\bar{\omega}} U_{\psi}^{-1} \bar{\psi}_{\bar{\omega}} \psi_{\bar{\omega}}
+
\frac{i}{\sqrt{\beta}} \Int{\bar{\omega} \omega} 
\big(
\bar{c}_{\bar{\omega}+\omega} \bar{d}_{\omega} \psi_{\bar{\omega}} -
d_{\omega} c_{\bar{\omega}+\omega} \bar{\psi}_{\bar{\omega}}
\big)
\EC
\nonumber \\
S_{\textrm{int}}
& = 
U \DimInt{\omega \nu \bar{\omega}} \bar{d}_{\omega} d_{\nu} \bar{c}_{\bar{\omega}+\nu} c_{\bar{\omega}+\omega}
\to
S_{\textrm{HS}} 
\EC
\quad
U_{\chi} + U_{\psi} = U
\ED
\label{eq:SHS}
\end{align}
Note that one can also set $U_{\chi}$ or $U_{\psi}$ to zero, 
such that one HS field effectively decouples from the system.
\begin{figure}[t]
\includegraphics[width=.485\textwidth]{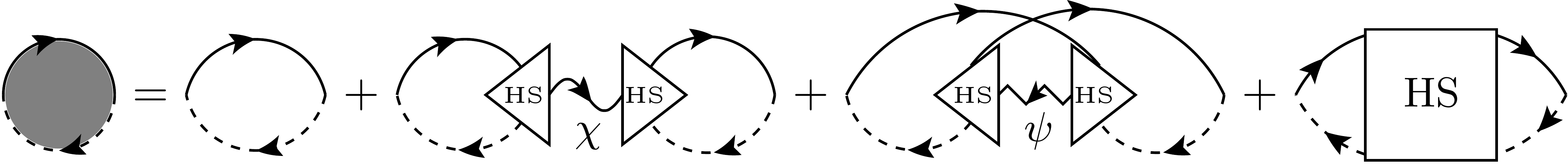}
\caption{%
Particle-hole susceptibility
after a HS transformation,
determined by HS three-point vertices and the four-point vertex
$\Gamma^{(4)}_{\textrm{HS}}$ (white square).
Wavy and zig-zag lines denote dressed
bosonic propagators.
Both three-point vertices
$\Gamma^{\bar{c} d \chi}$
and 
$\Gamma^{\bar{c} \bar{d} \psi}/i$
are depicted by a triangle
and can be distinguished by the attached bosonic line.%
}
\label{fig:corr_vertices}
\end{figure}
The more general relation between the particle-hole susceptibility and
1PI vertices in the presence of bosonic fields [cf.~Eq.~(6.92) of Ref.~\onlinecite{Kopietz2010}] 
is illustrated in \FR{fig:corr_vertices}.
Three-point vertices (denoted by triangles) 
and full bosonic propagators (wavy and zig-zag line)
contribute to the correlation function.
This proves beneficial in terms of computational effort as, next to the bosonic self-energies, the
three-point vertices
$\Gamma^{\bar{c} d \chi}_{\omega, \omega-\bar{\omega}, \bar{\omega}}$ and
$\Gamma^{\bar{c} \bar{d} \psi}_{\omega, \bar{\omega}-\omega, \bar{\omega}}/i$ 
(with initial condition unity)
contain less arguments
compared to the four-point vertex.
However, in \FR{fig:corr_vertices}, we see that
the particle-hole susceptibility is still directly affected
by the fermionic four-point vertex
(which is one-particle-irreducible in fermionic as well as bosonic lines).
The second and third summand on the r.h.s.\ take
the role of a four-point vertex reducible $\chi$ and $\psi$ lines, respectively,
and the actual four-point vertex still covers all contributions
irreducible in these lines.
Although the HS transformation by construction ensures that the 
four-point vertex does not contribute
to first order, it does comprise indispensable diagrams 
starting at second order in the interaction.
In \FR{fig:HSgamma4}(a), we show the simplest diagrams 
of $\Gamma^{(4)}_{\textrm{HS}}$
after the transformation, which now start at second order in $U$.
The lowest-order contributions to these diagrams,
obtained by using bare bosonic propagators,
represent the second-order ladder [with weight $U_{\psi}^2=(G^{\psi}_0)^2$] 
and second-order crossed diagram [with weight $U_{\chi}^2=(G^{\chi}_0)^2$],
known from \FR{fig:corr2nd} [cf.~\FsR{fig:CorrVertex_ParquetOrders}(b)
and \FRn{fig:Parquet3rdOrder}].
The main contributions of the exchange ($\chi$) and 
pairing ($\psi$) boson in \FR{fig:corr_vertices}
are reducible in the antiparallel and parallel (two-particle) channels, respectively.
Correspondingly, the lowest-order diagrams of $\Gamma^{(4)}_{\textrm{HS}}$
in \FR{fig:HSgamma4}(a) built from $\chi$ and $\psi$ lines
are reducible in the complementary channels, i.e., in parallel
and antiparallel (fermionic) lines, respectively.
However, starting at fourth order in the interaction,
also four-point-vertex diagrams with $\chi$ lines reducible
in the antiparallel channel exist, as is demonstrated in \FR{fig:HSgamma4}(b)
and analogously occurs with $\psi$ lines in the parallel channel.
In fact, the diagrams in \FR{fig:HSgamma4}(a) can be used as building blocks
that replace the bare interaction in the 
original parquet diagrams [cf.~\FR{fig:Parquet3rdOrder}]
to construct diagrams of $\Gamma^{(4)}_{\textrm{HS}}$.
Yet, this still covers only a fraction of the possible diagrams.
We conclude that obtaining the full weight for higher-order parquet contributions
to $\Pi$ via the relation in \FR{fig:corr_vertices}
requires a complicated, parquet-like resummation of diagrams containing
fermionic and bosonic lines in the four-point vertex.
\begin{figure}[t]
\includegraphics[width=0.22\textwidth]{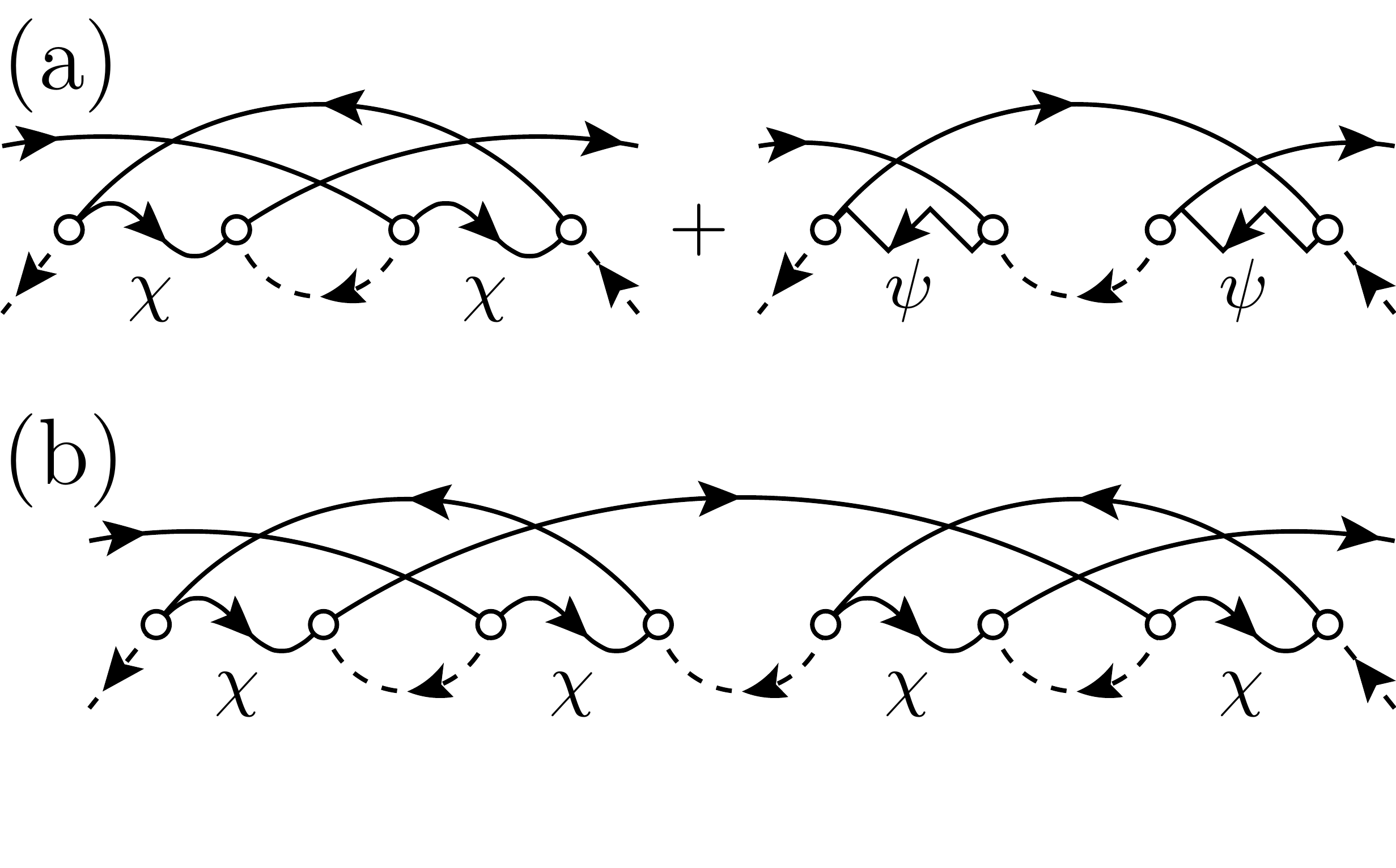}
\quad
\includegraphics[width=0.235\textwidth]{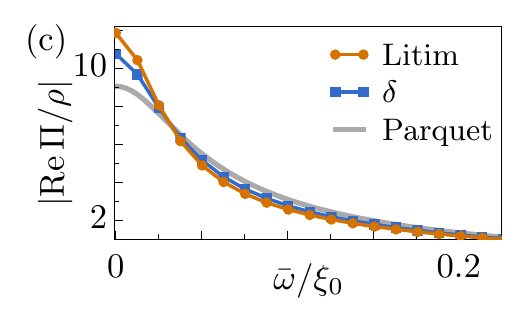}
\caption{%
(Color online)
(a)
After a HS transformation, $\Gamma^{(4)}_{\textrm{HS}}$
contributes with the above diagrams, starting at second order,
where white circles denote the bare three-point vertices, equal to unity.
It is therefore needed to produce all parquet graphs in the correlator.
(b)
Whereas diagrams of the four-point vertex in (a) that are
built from $\chi$ lines are irreducible
in the (corresponding) antiparallel channel,
starting at fourth order, diagrams with $\chi$ lines
that are reducible in antiparallel (fermionic) lines occur, too.
(c)
Particle-hole susceptibility
$\Pi$ computed via the relation in \FR{fig:corr_vertices} without $\Gamma^{(4)}_{\textrm{HS}}$,
where bosonic self-energies and three-point vertices are obtained 
from the truncated fRG flow \ERn{eq:HS_Sigma_flow}, \ERn{eq:HS_Gamma3_flow}, 
and the interaction strength is divided equally 
between both channels, $U_{\chi} = U/2 = U_{\psi}$.%
}
\label{fig:HSgamma4}
\end{figure}
The flow equations for the HS self-energies and three-point vertices
can be deduced from the fundamental flow equation \ERn{eq:gammaflow}.
When neglecting four-point and higher vertices,
they take a form which has already been 
given in Eqs.~(44), (45) of Ref.~\onlinecite{Lange2015}.
We repeat them here for the sake of completeness
and later purposes. The flow of the self-energies is given by
\begin{subequations}
\label{eq:HS_Sigma_flow}
\begin{align}
\partial_{\Lambda} \Pi^{\chi}_{\Lambda,\, \bar{\omega}} 
& = 
\DimInt{\omega}  
S^d_{\Lambda,\, \omega}
G^c_{\bar{\omega}+\omega}
\big( \Gamma^{\bar{c} d \chi}_{\Lambda,\, \bar{\omega}+\omega, \omega, \bar{\omega}} \big)^2
\EC
\\
\partial_{\Lambda} \Pi^{\psi}_{\Lambda,\, \bar{\omega}} 
& = 
\DimInt{\omega}  
S^d_{\Lambda,\, \omega}
G^c_{\bar{\omega}-\omega}
\big(
\Gamma^{\bar{c} \bar{d} \psi}_{\Lambda,\, \bar{\omega}-\omega, \omega, \bar{\omega}} /i
\big)^2
\ED
\end{align}
\end{subequations}
For the three-point vertices, one obtains
\begin{subequations}
\label{eq:HS_Gamma3_flow}
\begin{align}
\partial_{\Lambda} & \Gamma^{\bar{c} d \chi}_{\Lambda,\, \omega, \omega-\bar{\omega}, \bar{\omega}} 
=
\DimInt{\omega'}  
S^d_{\Lambda,\, \omega'}
\Gamma^{\bar{c} d \chi}_{\Lambda,\, \bar{\omega}+\omega', \omega', \bar{\omega}}
G^c_{\bar{\omega}+\omega'}
\nonumber \\ & \ \times
\Gamma^{\bar{c} \bar{d} \psi}_{\Lambda,\, \bar{\omega}+\omega', \omega-\bar{\omega}, \omega+\omega'} /i \,
G^{\psi}_{\Lambda,\, \omega+\omega'}
\Gamma^{\bar{c} \bar{d} \psi}_{\Lambda,\, \omega, \omega', \omega+\omega'}/i
\EC
\\
\partial_{\Lambda} & \Gamma^{\bar{c} \bar{d} \psi}_{\Lambda,\, \omega, \bar{\omega}-\omega, \bar{\omega}} /i
=
\DimInt{\omega'}  
S^d_{\Lambda,\, \omega'}
\Gamma^{\bar{c} \bar{d} \psi}_{\Lambda,\, \bar{\omega}-\omega', \omega', \bar{\omega}}/i \,
G^c_{\bar{\omega}-\omega'}
\nonumber \\ & \ \times
\Gamma^{\bar{c} d \chi}_{\Lambda,\, \bar{\omega}-\omega', \bar{\omega}-\omega, \omega-\omega'}
G^{\chi}_{\Lambda,\, \omega-\omega'}
\Gamma^{\bar{c} d \chi }_{\Lambda,\, \omega, \omega', \omega-\omega'}
\ED
\end{align}
\end{subequations}
To gauge the importance of the HS four-point vertex,
we have numerically solved the fRG flow in the HS-transformed system
[\EsR{eq:HS_Sigma_flow}, \ERn{eq:HS_Gamma3_flow}].
The resulting particle-hole susceptibility shown
in \FR{fig:HSgamma4}(c),
which is computed using the relation 
of \FR{fig:corr_vertices} without $\Gamma^{(4)}_{\textrm{HS}}$, 
shows much stronger deviations from the leading log result
than \FR{fig:Corr4}(a), which was obtained using only $\Gamma^{(4)}$.
This provides additional, numerical evidence
that a HS transformation does not save us 
from having to calculate the fermionic four-point vertex.
\section{Flowing susceptibility}
\label{sec:self-energy_flow}
An alternative approach to calculating the particle-hole susceptibility from renormalized 1PI vertices
is based on the identification of $\Pi$ as a bosonic self-energy.
In \ER{eq:phsuscep-photon}, we have shown how $\Pi$
is obtained from the self-energy of a rescaled photon field in the limit of its propagator
(containing the dipole matrix element) going to zero.
Flow equations for the photon self-energy without
internal photon propagation thus describe the flow
of the particle-hole susceptibility.
It should be noted that this appears natural given the interpretation of polariton physics,
but can also be seen as a mere computational trick
in order to directly include a susceptibility in the fRG flow.
In this section,
we consider the flow
of the photon self-energy in different
levels of truncation
and comment on the related publication by Lange et al.\ \cite{Lange2015}. 
\subsection{Dynamic four-point vertex -- numerical solution}
\begin{figure}[t]
\center
\includegraphics[width=.485\textwidth]{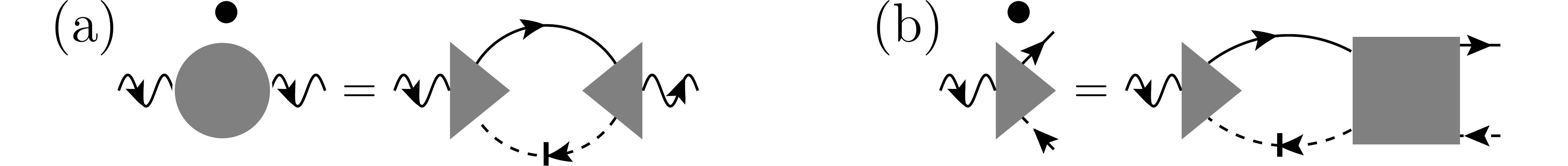}
\caption{%
Truncated flow equations for (a) the photon self-energy
$\Pi$ (depicted as circle)
and (b) the photon three-point vertex 
$\Gamma^{\bar{c} d \gamma}$ (depicted as triangle),
where the contributions of $\Gamma^{\bar{d} d \gamma \bar{\gamma}}$ [\ER{eq:photon_2flow}]
and $\Gamma^{\bar{c} d \gamma \bar{d} d}$ [\ER{eq:photon_3flow}] are neglected.
External (rapidly oscillating) wavy lines denote amputated photon legs.
Note that the truncated flow of the four-point vertex 
$\Gamma^{\bar{d} c \bar{c} d}$ is still given 
by \FR{fig:gamma4flow}(a).}
\label{fig:photon_flow}
\end{figure}
In the extended theory of the light-matter (photon and fermion) system,
we derive from the fundamental flow equation \ERn{eq:gammaflow}
the flow of the photon self-energy and three-point vertex:
\begin{subequations}
\label{eq:photon_flow}
\begin{align}
\partial_{\Lambda} \Pi_{\Lambda,\, \bar{\omega}} 
= 
\DimInt{\omega}  
&
S^d_{\Lambda,\, \omega}
\big[
G^c_{\bar{\omega}+\omega}
\big(
\Gamma^{\bar{c} d \gamma}_{\Lambda,\, \bar{\omega}+\omega, \omega, \bar{\omega}}
\big)^2
+
\Gamma^{\gamma \bar{\gamma} \bar{d} d}_{\Lambda,\, \bar{\omega}, \bar{\omega}, \omega, \omega}
\big]
\EC
\label{eq:photon_2flow}
\\
\partial_{\Lambda} 
\Gamma^{\bar{c} d \gamma}_{\Lambda,\, \omega, \omega-\bar{\omega}, \bar{\omega}}
&
=  
\DimInt{\omega'}  
S^d_{\Lambda,\, \omega'}
\big(
\Gamma^{\bar{c} d \gamma}_{\Lambda,\, \bar{\omega}+\omega', \omega', \bar{\omega}}
G^c_{\bar{\omega}+\omega'}
\nonumber \\
& 
\!\!\!\!\!\!\!\!\!\!
\times
\Gamma^{\bar{d} c \bar{c} d}_{\Lambda,\,
{\omega', \bar{\omega}+\omega'}, \omega, \omega-\bar{\omega}}
+
\Gamma^{\bar{c} d \gamma \bar{d} d}_%
{\Lambda,\, \omega, \omega-\bar{\omega}, \bar{\omega}, \omega', \omega'}
\big)
\ED
\label{eq:photon_3flow}
\end{align}
\end{subequations}
The flow of $\Gamma^{\bar{d} c \bar{c} d}$,
relevant for the second differential equation \ERn{eq:photon_3flow},
is still given by \ER{eq:gamma4flow}. In general,
three-point vertices connecting bosons and fermions 
would alter the flow of
$\Gamma^{\bar{d} c \bar{c} d}$,
but in the limit $G_0^{\gamma} \to 0$ these terms drop out.
Similarly, in the absence of propagating photons,
one finds that the (interband) flow of $\Gamma^{\gamma \bar{\gamma} \bar{d} d}$ is only 
determined by five- and six-point vertices.
At our level of truncation $\Gamma_{\Lambda}^{(n>4)} = \Gamma_{\Lambda_i}^{(n>4)} = 0$,
it is therefore consistent to set 
$\Gamma^{\gamma \bar{\gamma} \bar{d} d}_{\Lambda} = \Gamma^{\gamma \bar{\gamma} \bar{d} d}_{\Lambda_i} = 0$
alongside 
$\Gamma^{\bar{c} d \gamma \bar{d} d}_{\Lambda} = \Gamma^{\bar{c} d \gamma \bar{d} d}_{\Lambda_i} = 0$.
The resulting simplified flow
is illustrated in \FR{fig:photon_flow}.
Note that the diagrammatic expansion of 
the three-point vertex
$\Gamma^{\bar{c} d \gamma}$
is immediately deduced from the Schwinger-Dyson equation 
[cf., e.g., Fig.\ 11.6(b) of Ref.~\onlinecite{Kopietz2010}]
shown in \FR{fig:sd}(a).
As a consequence of truncation,
the connection between $\Pi$ and $\Gamma^{\bar{d} c \bar{c} d}$ 
generated by the flow (via $\Gamma^{\bar{c} d \gamma}$, cf.~\FR{fig:photon_flow}) 
violates the basic relation between susceptibility 
and four-point vertex
that was given in \ER{eq:phsuscep-gamma4}.
This is, however, intended in order to
obtain new resummations, 
given an approximate four-point vertex,
from the explicit photon flow.
\begin{figure}[t]
\center
\includegraphics[width=.485\textwidth]{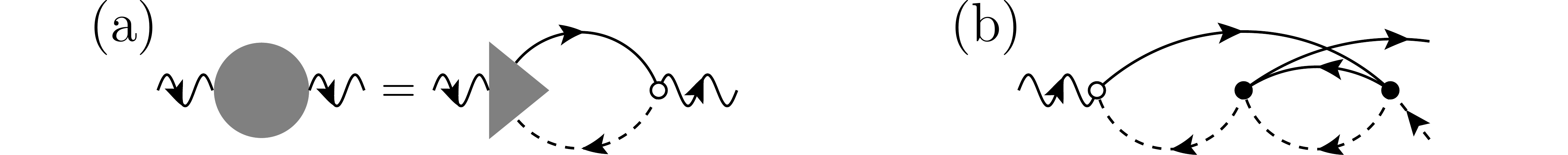}
\caption{%
(a)
Schwinger-Dyson equation between photon self-energy and three-point vertex,
where the small white circle denotes a bare photon three-point vertex, equal to unity.
(b) Second-order diagram of the three-point vertex, which [according to (a)]
is responsible for the crossed diagram
in the photon self-energy, viz., the
particle-hole susceptibility (cf.~\FR{fig:corr2nd}).%
}
\label{fig:sd}
\end{figure}
The numerical solution of the triple set of flow equations for
$\Pi$, $\Gamma^{\bar{c} d \gamma}$ (\FR{fig:photon_flow})
and $\Gamma^{\bar{d} c \bar{c} d}$ [\FR{fig:gamma4flow}(a)]
results in the 
particle-hole susceptibility
shown in \FR{fig:Corr4}(b).
The agreement between the numerical solution and the parquet formula
is worse compared to \FR{fig:Corr4}(a),
where only $\Gamma^{(4)}$ was used 
to compute $\Pi$.
The reason is that the additional flow equations in \FR{fig:photon_flow}
exclusively contain antiparallel $S^d$-$G^c$ lines.
They therefore induce an imbalance between the two-particle channels
and neglect important contributions of diagrams with parallel lines.
This begins with the crossed diagram at second order (cf.~\FR{fig:corr2nd}),
which is known \cite{Mahan1967}
to give a positive contribution 
to the particle-hole susceptibility
and thus reduce the infrared divergence.
So far, the more complicated way
to generate the particle-hole susceptibility from the four-point vertex,
namely the additional photon flow [\ER{eq:photon_flow}, \FR{fig:photon_flow}]
instead of the direct relation 
[\ER{eq:phsuscep-gamma4}, \FR{fig:CorrVertex_ParquetOrders}(b)],
has led to worse agreement with the leading log formula.
It is an underlying expectation of (vertex-expanded) fRG
that, by incorporating more vertices in the flow,
one improves the results,
coming closer to the exact, infinite hierarchy of flow equations
and having agreement with higher orders in perturbation theory.
By contrast, in the next section, we show 
that if we approximate $\Gamma^{\bar{d} c \bar{c} d}$
in the simplest fashion possible---namely by the bare vertex---%
we actually reproduce the
precise leading log result.
\subsection{Static four-point vertex -- analytic solution}
\label{sec:analyticsolution}
The enormous simplification of using 
the bare four-point vertex throughout the flow has hardly any justification.
Yet, we will show that, with this simplification, the flow equations can 
be solved analytically to yield the parquet result without further approximations.
This demonstrates that one cannot judge about the content of the diagrammatic resummation
solely based on the final result for a specific quantity.
We will first present a purely algebraic derivation of the leading log formula
for the particle-hole susceptibility
and then illustrate the steps
to diagrammatically understand the underlying structure.

%
Let us adopt a harsh but concise truncation of the flow equations:
we keep all 1PI vertices starting from the four-point vertex at their initial value.
The only (interband) contribution with a non-vanishing value at $\Lambda_i$ is 
the fermionic four-point vertex $\Gamma^{\bar{d} c \bar{c} d}_{\Lambda}$, which thus remains
equal to $-U$ throughout the flow.
The simplified flow equations [cf.~\ER{eq:photon_flow}] then read
\begin{subequations}
\label{eq:trivflow}
\begin{align}
\partial_{\Lambda} \Pi_{\Lambda,\, \bar{\omega}}
& =
\DimInt{\omega}
S^d_{\Lambda,\, \omega} 
G^c_{\bar{\omega}+\omega}
\big(
\Gamma^{\bar{c} d \gamma}_{\Lambda,\, \bar{\omega}+\omega, \omega, \bar{\omega}}
\big)^2
\EC
\label{eq:trivflow2}
\\
\partial_{\Lambda}
\Gamma^{\bar{c} d \gamma}_{\Lambda,\, \omega, \omega-\bar{\omega}, \bar{\omega}}
& =
- U
\DimInt{\omega'}  
S^d_{\Lambda,\, \omega'}
G^c_{\bar{\omega}+\omega'}
\Gamma^{\bar{c} d \gamma}_{\Lambda,\, \bar{\omega}+\omega', \omega', \bar{\omega}}
\ED
\label{eq:trivflow3}
\end{align}
\end{subequations}
The important observation is that the first derivative (and consequently any higher 
derivative) 
of $\Gamma^{\bar{c} d \gamma}_{\Lambda}$ is independent of $\omega$, i.e.,
completely independent of the first argument. 
(The second argument is fixed by conservation, anyway.)
Since also the initial condition is independent of the first argument, 
the vertex only depends on $\bar{\omega}$, but not on $\omega$, for all scales.
(This is a consequence of our truncation as 
diagrams of $\Gamma^{\bar{c} d \gamma}_{\Lambda}$ 
such as the one in \FR{fig:sd}(b),
corresponding to the crossed diagram in the particle-hole susceptibility,
do depend on the fermionic frequencies.)
Since $\Gamma^{\bar{c} d \gamma}_{\Lambda}$ is independent of $\omega$,
the differential equations \ERn{eq:trivflow}
can be dramatically simplified: Using the definition
$g_{\Lambda,\, \bar{\omega}}
=
\big(
\Gamma^{\bar{c} d \gamma}_{\Lambda,\, \cdot, \cdot, \bar{\omega}}
\big)^2$, we get
\begin{subequations}
\label{eq:trivsimplflow}
\begin{align}
\partial_{\Lambda} g_{\Lambda,\, \bar{\omega}}
& =
- 2U
g_{\Lambda,\, \bar{\omega}}
\DimInt{\omega}
S^d_{\Lambda,\, \omega} 
G^c_{\bar{\omega}+\omega}
\EC
\label{eq:trivsimplflow3}
\\
\partial_{\Lambda} 
\Pi_{\Lambda,\, \bar{\omega}}
& =
g_{\Lambda,\, \bar{\omega}}
\DimInt{\omega}
S^d_{\Lambda,\, \omega} 
G^c_{\bar{\omega}+\omega}
=
- \frac{1}{2U} \partial_{\Lambda} g_{\Lambda,\, \bar{\omega}}
\ED
\label{eq:trivsimplflow2}
\end{align}
\end{subequations}
Evidently, $g_{\Lambda,\, \bar{\omega}}$ is given by an exponential 
of an auxiliary function $f_{\Lambda,\, \bar{\omega}}$,
\begin{equation}
g_{\Lambda,\, \bar{\omega}} = 
g_{\Lambda_i,\, \bar{\omega}} e^{-2uf_{\Lambda,\, \bar{\omega}} }
\EC
\ 
f_{\Lambda,\, \bar{\omega}}
=
\int_{\Lambda_i}^{\Lambda} \! \Dif{\Lambda'} \!
\! \!
\DimInt{\omega}
S^d_{\Lambda',\, \omega} G^c_{\bar{\omega}+\omega}
/ \rho
\EC
\end{equation}
and the self-energy becomes
\begin{equation}
\Pi_{\Lambda,\, \bar{\omega}}
=
\Pi_{\Lambda_i,\, \bar{\omega}}
-
\frac{g_{\Lambda_i,\, \bar{\omega}}}{2U}
\big[
e^{
-2u
f_{\Lambda,\, \bar{\omega}}
}
-
1
\big]
\ED
\end{equation}
Inserting the boundary conditions
$\Pi_{\Lambda_i}=0$ and
$g_{\Lambda_i}=1$,
when $\Lambda$ flows from $\infty$ to $0$, 
we get
\begin{equation}
\Pi_{\bar{\omega}}
=
\frac{1}{2U}
\big[
1
-
e^{
-2u
f_{\bar{\omega}}
}
\big]
\EC
\quad
f_{\bar{\omega}} =
\int_{\infty}^0 \Dif{\Lambda} \!\!
\DimInt{\omega}
S^d_{\Lambda,\, \omega} G^c_{\bar{\omega}+\omega}/ \rho
\ED
\label{eq:PiF}
\end{equation}
%

%
%

%
So far, fermionic self-energies have not been neglected, yet.
However, for the X-ray-edge singularity,
we can use $S^d_{\Lambda} = \partial_{\Lambda} G^d_{\Lambda}$ 
and the $\Lambda$-integration becomes trivial. 
Using the bare bubble, computed in \AR, \ER{eq:app:barebubble},
we arrive at the remarkable conclusion that our harsh truncation
directly yields the leading log result:
\begin{subequations}
\label{eq:phbubble_analyticparquet}
\begin{align}
f_{\bar{\omega}}
& =
\DimInt{\omega}
G^d_{\omega} G^c_{\bar{\omega}+\omega}
/ \rho
=
\ln \Big( \frac{i\bar{\omega} +\xi_d}{-\xi_0} \Big)
\EC
\label{eq:phbubble}
\\
\Pi_{\bar{\omega}}
& =
\frac{\rho}{2u}
\bigg[
1 -
\Big(
\frac{i\bar{\omega} +\xi_d}{-\xi_0}
\Big)
^{-2u}
\bigg]
\ED
\label{eq:analyticparquet}
\end{align}
\end{subequations}
How is this possible? We have argued above that,
in the combined, truncated system of flow equations for
$\Gamma^{\bar{d} c \bar{c} d}$ and photon quantities,
a large class of parquet contributions is missed by the 
approximate flow
due to a mistreatment of parallel bubbles.
We will now show diagrammatically why the parquet result
could nevertheless be obtained and will find that
this is only possible for the X-ray-edge singularity.
The diagrammatic solution of the simplified flow makes extensive
use of the property that ladder diagrams factorize into a sequence of
(particle-hole) bubbles
and that,
with leading log accuracy,
we can ignore fermionic self-energies and use $S^d = \partial_{\Lambda}G^d_0$.
If we use the bare four-point vertex
in the flow of the three-point vertex [\FR{fig:photon_flow}(b)],
we obtain the flow equation shown in \FR{fig:triv_g3_flow}(a),
which interrelates contributions to $\Gamma^{\bar{c} d \gamma}$ from subsequent orders.
Due to factorization, the solution to this flow equation 
can be expressed diagrammatically as a three-point vertex 
which, at order $n$, consists of $n$ consecutive
particle-hole bubbles multiplied by a
prefactor $1/n!$ [\FR{fig:triv_g3_flow}(b)]. 
The simple ladder structure is directly related to the fact that
$\Gamma^{\bar{c} d \gamma}_{\Lambda,\, \omega, \omega-\bar{\omega}, \bar{\omega}}$
is independent of $\omega$.
\begin{figure}[t]
\includegraphics[width=.48\textwidth]{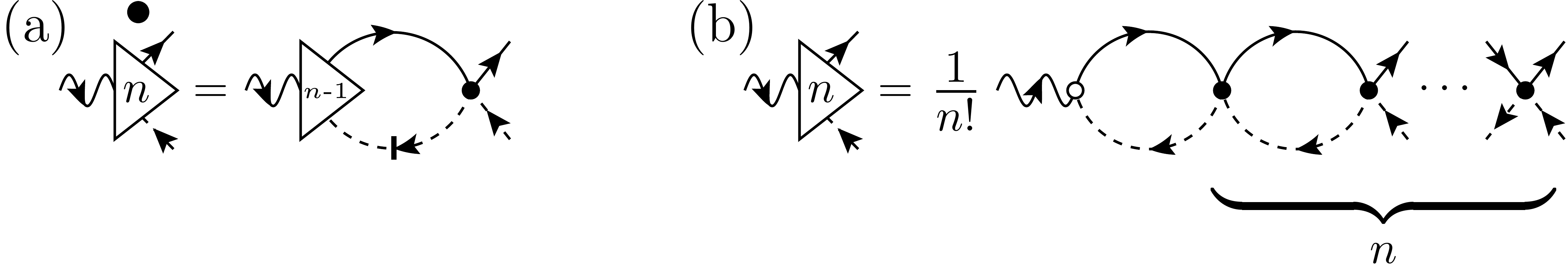}
\caption{%
(a) Flow equation for an approximate $\Gamma^{\bar{c} d \gamma}$ (at order $n$)
when $\Gamma^{\bar{d} c \bar{c} d}$ is reduced to its bare part
[cf.~\FR{fig:photon_flow}(b)].
(b) Its solution, given by sequence of bubbles
with a prefactor $1/n!$, a bare photon three-point vertex (equal to unity)
and $n$ bare electronic interaction vertices.
}
\label{fig:triv_g3_flow}
\end{figure}
Inserting this three-point vertex in the flow equation of the 
photon self-energy [\FR{fig:photon_flow}(a)],
we get, at order $n$, a sequence of $n+1$ bubbles
with one single-scale propagator (cf.~\FR{fig:pi_proof}).
Again using factorization, this is a fraction [$1/(n+1)$] 
of the derivative of the whole ladder diagram.
By computing the sum 
$\sum_{m=0}^{n} 1/[m!(n-m)!] = 2^n/n!$ in \FR{fig:pi_proof},
one ends up with a proportionality relation (at arbitrary order $n$)
between the derivative of the self-energy, $\partial_{\Lambda} \Pi^{(n)}$,
and the derivative of a ladder-diagram, $\partial_{\Lambda} \Pi^{\textrm{L}(n)}$.
As these quantities also agree at the initial scale (both vanish when $G^d = 0$),
we extract an equality at all scales.
Using the bare bubble as in \ER{eq:phbubble}, we get
\begin{equation}
\Pi^{(n)}_{\bar{\omega}}
=
\frac{2^n}{(n+1)!} 
\Pi^{\textrm{L}(n)}_{\bar{\omega}}
\EC \quad
\Pi^{\textrm{L}(n)}_{\bar{\omega}}
=
(-U)^n ( \rho f_{\bar{\omega}} )^{n+1}
\ED
\label{eq:piladder}
\end{equation}
It remains to sum all orders $\Pi^{(n)}_{\bar{\omega}}$,
i.e., sum all ladder diagrams with the appropriate prefactor [cf.~\ER{eq:piladder}].
Indeed, we precisely reproduce the leading log result
\begin{align}
\Pi_{\bar{\omega}}
& =
\sum\limits_{n=0}^{\infty}
\Pi^{(n)}_{\bar{\omega}}
=
- \frac{1}{2U}
\sum\limits_{n=0}^{\infty}\
\frac{( -2 u f_{\bar{\omega}} )^{n+1}}{(n+1)!} 
\nonumber \\
& =
- \frac{\rho}{2u}
\Big(
e^{-2u f_{ \bar{\omega} } } - 1
\Big)
=
\frac{\rho}{2u}
\bigg[
1 - \Big( \frac{i\bar{\omega} + \xi_d}{-\xi_0} \Big)^{-2u}
\bigg]
\ED
\label{eq:graphparquet}
\end{align}
\begin{figure}[t]
\includegraphics[width=.48\textwidth]{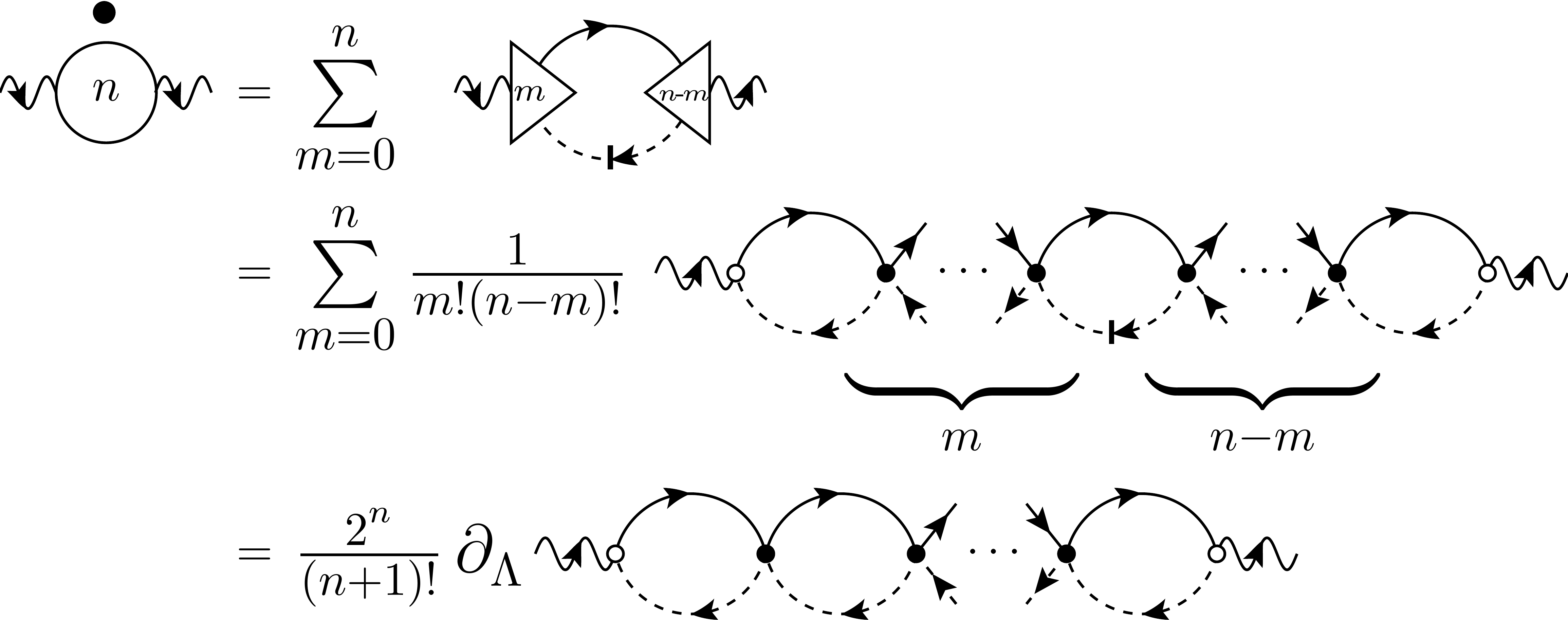}
\caption{%
Inserting the approximate $\Gamma^{\bar{c} d \gamma}$ from \FR{fig:triv_g3_flow}(b)
in the simplified flow of $\Pi$ [\FR{fig:photon_flow}(a)],
we obtain a proportionality relation between ladder diagrams
and the particle-hole susceptibility at arbitrary order $n$,
in exact agreement with the leading log result [cf.~\ER{eq:graphparquet}].%
}
\label{fig:pi_proof}
\end{figure}
We observe that \textit{only} ladder diagrams are generated by the flow
while crossed diagrams do not contribute at all.
However, the ladder diagrams come with prefactors,
such as $1/n!$ in \FR{fig:triv_g3_flow}(b) and $2^n/(n+1)!$ 
in \ER{eq:piladder}.
That the correct form of the particle-hole susceptibility
is obtained at every order is then possible due to
proportionality relations present in the X-ray-edge singularity,
such as $\Pi^{\textrm{L}(2)} = -3 \Pi^{\textrm{C}(2)}$
[cf.~\FR{fig:corr2nd}], as already shown by Mahan \cite{Mahan1967}
fifty years ago.
Yet, these relations only hold with logarithmic accuracy,
and in the more general Fermi-edge singularity, where
the assumption of an infinite hole mass is lifted,
they hold only in a very narrow parametric regime 
(namely for $m/m_h$ being exponentially
small in the coupling $u$) \cite{Pimenov2015,*Pimenov2017, Gavoret1969}.
For other problems, surely such relations 
will only hold, if at all, subject to further assumptions.
We therefore conclude that obtaining the exact first-order parquet result 
from a truncated fRG flow with a static four-point vertex
is only possible due to a fortuitous partial cancellation of diagrams,
specific to the X-ray-edge singularity.
\subsection{Comparison to a work by Lange et al.}
In a recent publication, Lange, Drukier, Sharma, and Kopietz \cite{Lange2015} (LDSK) 
have addressed the question of using fRG to tackle the X-ray-edge singularity. 
In fact, it is their paper which has drawn our attention to the problem at hand
and deeply inspired our approach.
LDSK, too, obtain the (first-order)
parquet formula for the particle-hole susceptibility
[our \ER{eq:parquet} and their Eq.~(54)]
and from this draw conclusions about the relation between 
parquet summations and fRG.
We hope that our analysis has further elucidated
the derivation of the analytic result
and added valuable arguments to the discussion
about fRG and parquet graphs.
Let us comment on some interesting points from LDSK's treatment in detail.
LDSK extract the particle-hole susceptibility from a bosonic self-energy ($\Pi^{\chi}$)
arising from a multi-channel Hubbard-Stratonovich (HS) transformation 
in the exchange ($\chi$, $U_{\chi}$) and pairing ($\psi$, $U_{\psi}$) channel.
They choose (i) equal weights in both channels, $U_{\chi}=U_{\psi}$,
while we will argue that only the choice $U_{\chi}=0$ allows the particle-hole
susceptibility to be extracted correctly from the $\chi$ self-energy. 
We will (ii) further show that, with the choice $U_{\chi}=0$,
one can avoid one of the approximations
made by LDSK, namely to take $u \ln(\xi_0 / |\bar{\omega}| ) \ll 1$.
We will (iii) comment on the similarity between our approximate
flow in the light-matter system and LDSK's
flow in the HS-transformed system and
demonstrate numerically that including the HS-bosonic self-energies
weakens the agreement with the parquet result.
Furthermore, LDSK use an approximation scheme where all frequency dependencies
are initially neglected and finally restored by stopping the RG flow
at a final value of $\Lambda_f = \bar{\omega}$. We will (iv) give
an argument, using the $\delta$ regulator, for why this scheme successfully leads to
the parquet result.
\begin{figure}[t]
\includegraphics[width=.48\textwidth]{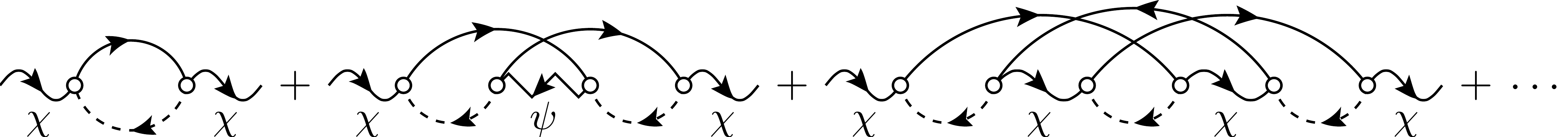}
\caption{%
Diagrams for the $\chi$ self-energy $\Pi^{\chi}$, expressed with bare three-point
vertices (small white circles), equal to unity.
At zeroth order in $U$, $\Pi^{\chi}$ is given by a bare particle-hole bubble;
the only first-order contribution arises from the second diagram using $G_0^{\psi} = -U_{\psi}$.
Starting at second order in the interaction, 
$\Pi^{\chi}$ contains diagrams with internal $\chi$ lines,
as in the third diagram above.%
}
\label{fig:chi_selfenergy}
\end{figure}
(i)
From the actions in \EsR{eq:Sgamma} and \ERn{eq:SHS},
it is clear that the HS field in the exchange channel, $\chi$, 
couples similarly to fermions as the photon field $\gamma$.
However, just as for the photon [cf.~\ER{eq:phsuscep-photon}],
it is crucial that the particle-hole susceptibility $\Pi$ be
fully represented by only the \textit{leading} part of
the $\chi$ self-energy $\Pi^{\chi}$, i.e.,
the part without internal $\chi$ propagation.
This is easily seen in terms of diagrams (\FR{fig:chi_selfenergy}):
$\Pi^{\chi}$ at zeroth order is given by a 
conduction-valence-band particle-hole bubble,
representing the zeroth-order contribution to $\Pi$.
At first order in the interaction, $\Pi^{\chi}$ is affected 
solely by $\psi$ propagation,
for an intermediate $\chi$ line would result in a reducible diagram.
Hence, for $\Pi^{\chi}$ to fully account for the 
first-order ladder diagram of $\Pi$, 
the bare $\psi$ propagator must have full weight,
$U_{\psi} = U$.
On the other hand, at second and higher orders, $\Pi^{\chi}$
contains irreducible diagrams with internal $\chi$ lines.
If one chose $U_{\chi}>0$, one would overcount these contributions
and not properly generate the second-order order contribution to $\Pi$.
Hence, the exact parquet graphs for $\Pi$ can only
be reproduced from $\Pi^{\chi}$ by using $U_{\psi} = U$ and $U_{\chi} = 0$.
(ii)
Interestingly enough, with the latter choice,
the approximate analytic approach of LDSK
can be simplified. LDSK use $U_{\chi}=U_{\psi}=U$ and arrive
at an integration of the frequency-independent, squared $\chi$ three-point vertex $g_l$
from a logarithmic scale parameter $l=0$ up to $l^* = \ln(\xi_0 / |\bar{\omega}| )$.
There, they approximate $\cosh(2ul)$ by unity [their Eq.~(52)], although 
$ul \ll 1$ holds no longer when $l$ reaches the upper integration limit,
since in the first-order parquet regime $ul^* = u \ln(\xi_0 / |\bar{\omega}| ) \lesssim 1$.
If one avoids this approximation and instead uses the
actual $g_l = e^{2ul}/\cosh(2ul)$ for the integral in LDSK's Eq.~(52),
one obtains
\begin{align}
\Pi^{\chi}_{\bar{\omega}} & = - \rho \int_0^{l^*} \Dif{l} \frac{e^{2ul}}{\cosh(2ul)} = 
- \frac{\rho}{2u} \ln \bigg( \frac{e^{4ul^*}+1}{2} \bigg)
\nonumber \\
& = 
- \rho l^* - \rho u {l^*}^2 + \mathcal{O}(u^3)
\EC
\end{align}
This contains no second-order term and thus deviates
already at second order in $U$ from the parquet result \ERn{eq:parquet}.
Note that, with $\xi_d=0$ (as chosen by LDSK), one can only
obtain the real part of the particle-hole susceptibility,
solely depending on $|\bar{\omega}|$ (cf.~\AR).
In this case, an expansion of \ER{eq:parquet} yields
\begin{align}
\textrm{Re } \Pi_{\bar{\omega}}\Big|_{\xi_d=0} & =
\frac{\rho}{2u} \Bigg[ 1 - \bigg( \frac{|\bar{\omega}|}{\xi_0} \bigg)^{-2u} \Bigg]
=
\frac{\rho}{2u} \big( 1 - e^{2ul^*} \big)
\nonumber \\
& =  
- \rho l^* - \rho u {l^*}^2 - \tfrac{2}{3} \rho u^2 {l^*}^3 + \mathcal{O}(u^3)
\ED
\label{eq:re_parquet}
\end{align}
The reason why performing the integral
more accurately leads to an incorrect result
is that the expression $g_l = e^{2ul}/\cosh(2ul)$
is inaccurate at second order, since it was 
obtained using $U_{\chi} \neq 0$.
(Consequently, $\Pi^{\chi}$ deviates from $\Pi$ starting at second order,
consistent with our diagrammatic argument above.)
If, instead, one uses $U_{\chi} = 0$ and $U_{\psi} = U$,
then Eq.~(49a) of LDSK naturally yields
$g_l = e^{2ul}$ instead of $g_l = e^{2ul}/\cosh(2ul)$,
so that the integration in their Eq.~(52) reads
\begin{align}
\Pi^{\chi}_{\bar{\omega}} & = -\rho \int_0^{l^*} \Dif{l} e^{2ul} = 
\frac{\rho}{2u} \big( 1 - e^{2ul^*} \big)
\end{align}
and precisely reproduces the result of \ER{eq:re_parquet}.
(iii)
If one sets $U_{\chi}=0$
in LDSK's flow equations (44), (45) 
[our \EsR{eq:HS_Sigma_flow}, \ERn{eq:HS_Gamma3_flow}],
the three-point vertex $\Gamma^{\bar{c} \bar{d} \psi}/i$ remains equal to unity,
since $G^{\chi}=0$ implies $\partial_{\Lambda} \Gamma^{\bar{c} \bar{d} \psi}=0$.
If one further omits bosonic self-energy reinsertions (as done by LDSK), 
one has $G^{\psi} = -U_{\psi} = -U$.
Hence, the resulting flow equations
for $\Pi^{\chi}$ and $\Gamma^{\bar{c} d \chi}$ reduce to exactly the form
of our \ER{eq:trivflow}
(replacing $\gamma$ by $\chi$).
As we have shown, this flow yields the leading log result
for the particle-hole susceptibility without further approximations.
Actual effects of the multi-channel HS transformation
become noticeable only if one actually includes
bosonic self-energies on the r.h.s.\ of the HS flow
[\EsR{eq:HS_Sigma_flow}, \ERn{eq:HS_Gamma3_flow}].
Figure \FRn{fig:HSselfenergy} shows (a)
that, in the case of $U_{\chi}=0$, $U_{\psi}=U$,
this spoils the agreement with the leading log result
and (b) the strikingly different outcome when
using  $U_{\chi} = U_{\psi} = U/2$.
In the latter case, 
$\Gamma^{\bar{c} \bar{d} \psi}$ contributes non-trivially,
and the result is more similar to that of the leading log formula with $U/2$,
since the effect of using $U_{\chi}>0$ enters only at second and higher orders
(cf.~\FR{fig:chi_selfenergy}).
We conclude that a (multi-channel) HS transformation has no advantage 
over the version advocated in \SR{sec:self-energy_flow} of this work, 
based on a flowing susceptibility in the fermionic system.
\begin{figure}[t]
\includegraphics[width=.48\textwidth]{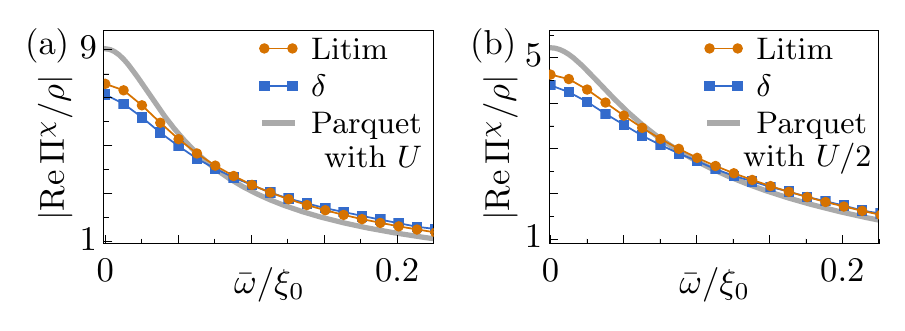}
\caption{%
(Color online)
Self-energy of $\chi$, $\Pi^{\chi}$, as obtained from the flow
in the HS-transformed system (neglecting $\Gamma^{(n > 3)}$)
[cf.~\EsR{eq:HS_Sigma_flow}, \ERn{eq:HS_Gamma3_flow}].
The interaction strength is divided according to
(a) $U_{\chi}=0$, $U_{\psi}=U$ and
(b) $U_{\chi}=U/2=U_{\psi}$. 
Since $\chi$ propagation affects $\Pi^{\chi}$ only starting at second order
(cf.~\FR{fig:chi_selfenergy}),
the result in (b) is more similar to the leading log formula with $U/2$.%
}
\label{fig:HSselfenergy}
\end{figure}
(iv)
In their analytic solution of the flow, LDSK use an approximation scheme
where frequency dependencies in all 1PI vertices were omitted initially.
Viewing this as a low-energy approximation, 
they let $\Lambda$ flow 
from $\xi_0$ to $\bar{\omega}$ instead of the expected range $\infty$ to $0$.
From another perspective, this integration range for $\Lambda$
can be obtained by computing the ``single-scale'' bubble [\ER{eq:ssbubble}]
with the $\delta$ regulator.
As explained above, LDSK's system of flow equations
with $U_{\chi}=0$ and $G^{\psi} = -U$ can be directly
related to our photon flow in \ER{eq:trivflow}.
We have shown that the $\bar{\omega}$-dependence
enters only in the (integrated) single-scale bubble 
[$f_{\bar{\omega}}$ in \ER{eq:PiF}],
which can also be integrated \textit{first} w.r.t.\ frequency
and \textit{then} w.r.t.\ $\Lambda$.
Making use of the $\delta$ regulator, $\xi_d=0$ (such that $|\bar{\omega}| \ll \xi_0$),
and the (simplified) local $c$ propagator [\ER{eq:gclocal}],
one readily obtains
\begin{align}
\DimInt{\omega}
S^d_{\Lambda,\, \omega} G^c_{\bar{\omega}+\omega} / \rho
& = 
\int_{-\xi_0-\bar{\omega}}^{\xi_0-\bar{\omega}}
\Dif{\omega} \textrm{sgn}(\bar{\omega}+\omega)
\frac{ \delta \big( |\omega|-\Lambda \big) }{2 \omega}
\nonumber \\
& \approx
\int_{-\xi_0}^{\xi_0}
\Dif{\omega} \textrm{sgn}(\bar{\omega}+\omega)
\frac{ \delta \big( |\omega|-\Lambda \big) }{2 \omega}
\nonumber \\
& 
=
\Theta \big( \xi_0 - \Lambda \big)
\sum_{\omega = \pm \Lambda}
\frac{\textrm{sgn}(\bar{\omega}+\omega) }{2\omega}
\nonumber \\
& 
=
\frac{ \Theta \big( \xi_0 - \Lambda \big) \Theta \big( \Lambda - |\bar{\omega}| \big)  }{\Lambda}
\ED
\label{eq:ssbubble}
\end{align}
Using this as a factor in the relevant flow equations,
similarly as in \ER{eq:trivsimplflow},
naturally restricts the integration range for $\Lambda$ precisely 
in the way chosen by LDSK.
\section{Conclusions}
\label{sec:conclusion}
We have analyzed the X-ray-edge 
(zero-dimensional Fermi-edge)
singularity---%
an instructive fermionic problem
with simplified diagrammatics
focused on two-particle quantities,
an analytic parquet and exact one-body solution.
Our goal was to use the functional renormalization group 
to achieve a partial resummation of diagrams, to
be compared to the (first-order) solution of the
parquet equations.
We compared results for the particle-hole susceptibility
with the leading log formula in terms of Matsubara frequencies
and examined the diagrammatic structure of the flow equations.
We found that 
different realizations of a truncated, one-loop
fRG flow do not fully generate the leading log diagrams.
Focusing on the flow of the fermionic four-point vertex $\Gamma^{(4)}$ first,
we argued that, in the fRG hierarchy, the parquet 
diagrams comprise (1PI and 2PI) vertices
of any order, and that these higher-order vertices,
obtained by cutting appropriate scale-dependent lines,
universally contribute to the flow.
Hence, simply truncating the fRG hierarchy of flow equations will always miss 
contributions to parquet graphs.
We further showed that a (multi-channel)
Hubbard-Stratonovich transformation
does not remedy this problem:
Although the transformation ensures that $\Gamma^{(4)}_{\textrm{HS}}$ does not contribute to the
particle-hole susceptibility $\Pi$ at first order,
it does contribute important, parquet diagrams to $\Pi$ starting at second order
in the interaction, which are lost 
when the four-point vertex is neglected.
As a different approach,
we included $\Pi$ in the fRG flow 
as a (leading contribution to a) photon self-energy
(i.e., as a flowing susceptibility).
We showed that
the relation between $\Gamma^{(4)}$ and $\Pi$
generated by truncated flow equations
systematically misses contributions from parallel bubbles.
However, in contrast to the underlying philosophy of fRG,
we found an \textit{improved} result for $\Pi$ when treating the four-point vertex
\textit{less} accurately.
In fact,
we analytically reproduced the 
leading log formula
using a truncated fRG flow that keeps four-point and higher vertices \textit{constant}.
We showed that, in this way, one effectively \textit{only} sums up ladder diagrams,
but with a set of prefactors that fortuitously 
turns out to precisely yield the correct form of $\Pi$.
This is possible thanks to proportionality relations
of ladder and crossed diagrams, which,
however, 
only hold with logarithmic accuracy
and are violated when extending the theory,
e.g., to a finite-mass valence-band description.
Our derivation of the (first-order) parquet result from 
a truncated fRG flow using a static four-point vertex 
is thus only possible due to a fortuitous partial
cancellation of diagrams specific to the
X-ray-edge singularity.
In related publications \cite{Kugler2017, Kugler2017a}, 
we show how the truncated flow equations \textit{can} actually
be extended to capture all parquet graphs.
This \textit{multiloop} fRG flow
simulates the effect of the six-point vertex
on parquet contributions and
iteratively completes the derivative of diagrams
in the flow equations of both four-point vertex
and self-energy.
\begin{acknowledgments}
We thank D.\ Pimenov, D.\ Schimmel, and L.\ Weidinger for
useful discussions and P.\ Kopietz for a helpful correspondence.
We acknowledge support by the Cluster of Excellence
Nanosystems Initiative Munich; F.B.K.\ acknowledges
funding from the research school IMPRS-QST.
\end{acknowledgments}

\appendix*
\section{Particle-hole bubble}
\label{appendix}
In this section, we explicitly compute the bare 
(interband) particle-hole bubble, needed
in \SR{sec:self-energy_flow}, \ER{eq:phbubble_analyticparquet}.
We also show that this bubble is discontinous w.r.t.\ 
the bandgap $-\xi_d$ at $\xi_d=0$.
Thus,
we choose $\xi_d$ suitably small
(cf.~\SR{sec:parquet})
but nonzero in our numerical calculations.
The bare bubble is given by the integral
\begin{align}
\Pi_{0, \bar{\omega}} & = \DimInt{\omega} G_0^c(\bar{\omega}+\omega) G_0^d(\omega)
\nonumber
\\
& = -i\pi \rho \DimInt{\omega} \frac{ \textrm{sgn}(\bar{\omega}+\omega) \Theta(\xi_0 - |\bar{\omega}+\omega|)}{i\omega-\xi_d}
\nonumber
\\
& = \Pi_{0, -\bar{\omega}}^*
\EC
\label{eq:app:pisym}
\end{align}
which we divide into three parts:
$\Pi_{0, \bar{\omega}} = I_1 + I_2 + I_3$.
We first consider $\bar{\omega}>0$,
revert to frequency integrals in 
the zero-temperature limit [cf.~\ER{eq:zerotemp}],
and obtain
\begin{subequations}
\label{eq:app:I}
\begin{align}
I_1 & = 
\frac{\rho}{2i}
\int_{\bar{\omega}}^{\xi_0-\bar{\omega}} 
\frac{\Dif{\omega}}{i\omega-\xi_d}
= 
\frac{\rho}{2}
\ln \Big( \frac{i\bar{\omega}-\xi_d}{i\xi_0-i\bar{\omega}-\xi_d} \Big)
\EC
\label{eq:app:I1} \\ 
I_2 & = 
\frac{\rho}{2i}
\int_{-\bar{\omega}}^{\bar{\omega}} 
\frac{\Dif{\omega}}{i\omega-\xi_d}
= 
\frac{\rho}{2}
\ln \Big( \frac{-i\bar{\omega}-\xi_d}{i\bar{\omega}-\xi_d} \Big)
\EC
\label{eq:app:I2log} \\ 
I_2 & = 
i\rho
\int_{0}^{\bar{\omega}} 
\Dif{\omega} \frac{\xi_d}{(\xi_d)^2+(\bar{\omega})^2}
= i\rho \arctan \Big( \frac{\bar{\omega}}{\xi_d} \Big)
\EC
\label{eq:app:I2} \\ 
I_3 & = 
\frac{i\rho}{2} \int_{-\xi_0-\bar{\omega}}^{-\bar{\omega}} 
\frac{\Dif{\omega}}{i\omega-\xi_d}
= 
\frac{\rho}{2}
\ln \Big( \frac{i\bar{\omega}+\xi_d}{i\xi_0+i\bar{\omega}+\xi_d} \Big)
\ED
\label{eq:app:I3}
\end{align}
\end{subequations}
In the form of \ER{eq:app:I2},
one can directly see that the integral $I_2$ is discontinuous w.r.t.\ $\xi_d$
at $\xi_d=0$. 
Essentially, the contribution from $I_2$ is 
needed to produce the correct phase in the susceptibility,
when summing $I_1$, $I_2$, and $I_3$.
Using the fact that, upon analytic continuation to real frequencies, one has
$|i\bar{\omega}+\xi_d| \to |\omega+\xi_d+i0^+| \ll \xi_0$, we obtain
the approximate form
\begin{align}
\Pi_{0, \bar{\omega}} & = 
\frac{1}{2}
 \ln \Big( \frac{i\bar{\omega}+\xi_d}{i\bar{\omega}+\xi_d-i\xi_0} \Big)
+
\frac{1}{2}
\ln \Big( \frac{i\bar{\omega}+\xi_d}{i\bar{\omega}+\xi_d+i\xi_0} \Big)
\nonumber
\\ & 
\approx
\frac{1}{2}
\ln \Big( \frac{i\bar{\omega}+\xi_d}{-i\xi_0} \Big)
+
\frac{1}{2}
\ln \Big( \frac{i\bar{\omega}+\xi_d}{i\xi_0} \Big)
\nonumber
\\ & 
=
\ln \Big( \frac{i\bar{\omega}+\xi_d}{-\xi_0} \Big)
\EC
\label{eq:app:barebubble}
\end{align}
which also holds for negative frequencies according to
the symmetry relation
$\Pi_{0, \bar{\omega}} = \Pi_{0, -\bar{\omega}}^{*}$.

If, instead, one sets $\xi_d=0$ in the first place, 
one in effect omits the contribution from $I_2$ [cf.~\ER{eq:app:I2}].
With the approximation $|i\bar{\omega}| \ll \xi_0$,
one then obtains from $I_1 + I_3$:
\begin{align}
\Pi_{0, \bar{\omega}}\Big|_{\xi_d=0, \,\bar{\omega}>0} & =
\frac{1}{2} \ln \Big( \frac{i\bar{\omega}}{i\xi_0 -i\bar{\omega}} \Big)
+
\frac{1}{2} \ln \Big( \frac{i\bar{\omega}}{i\xi_0 +i\bar{\omega}} \Big)
\nonumber
\\ & 
\approx
\ln \Big( \frac{\bar{\omega}}{\xi_0} \Big)
\ED
\end{align}
Reverting to positive and negative frequencies
via \ER{eq:app:pisym} again, we finally get
\begin{equation}
\Pi_{0, \bar{\omega}}\Big|_{\xi_d=0} 
= \ln \Big( \frac{|\bar{\omega}|}{\xi_0} \Big)
\ED
\end{equation}
Having set $\xi_d=0$,
one only obtains the real part of the particle-hole bubble,
solely depending on $|\bar{\omega}|$.
Moreover, in contrast to the real-frequency calculations of 
Roulet et.\ al \cite{Roulet1969}, who focus on the real part
and argue that the imaginary part can be reconstructed
by Kramers-Kronig relations,
this is not possible in the Matsubara framework,
where one does \textit{not} have such relations 
between $\textrm{Re } \Pi$ and $\textrm{Im } \Pi$. 
We conclude that one should therefore refrain from
setting $\xi_d=0$.

\bibliographystyle{apsrev4-1}
\bibliography{references}

\end{document}